\documentclass[
superscriptaddress,
aip,apl,
reprint,
twocolumn,
amsmath,amssymb,showpacs,
]{revtex4-2}
\usepackage{titlesec}
\usepackage{mathptmx}  
\usepackage{physics}
\usepackage{graphicx}
\usepackage{bm}
\usepackage{dsfont}
\usepackage{xcolor}
\definecolor{urlc}{RGB}{58,105,157}
\usepackage[
colorlinks=true,
urlcolor=urlc,
linkcolor=urlc,
citecolor=urlc
]{hyperref}

\titleformat{\section}{\large\sffamily\bfseries}{\thesection}{}{}
\begin{document}

\title{Experimental quantum simulation of superradiant phase transition beyond no-go theorem via antisqueezing}

\author{Xi Chen}
\thanks{These authors contributed equally to this work}
\affiliation{
Hefei National Laboratory for Physical Sciences at the Microscale and Department of Modern Physics, University of Science and Technology of China, Hefei 230026, China}
\affiliation{
CAS Key Laboratory of Microscale Magnetic Resonance, University of Science and Technology of China, Hefei 230026, China}
\affiliation{
Synergetic Innovation Center of Quantum Information and Quantum Physics, University of Science and Technology of China, Hefei 230026, China}

\author{Ze Wu}
\thanks{These authors contributed equally to this work}
\affiliation{
Hefei National Laboratory for Physical Sciences at the Microscale and Department of Modern Physics, University of Science and Technology of China, Hefei 230026, China}
\affiliation{
CAS Key Laboratory of Microscale Magnetic Resonance, University of Science and Technology of China, Hefei 230026, China}
\affiliation{
Synergetic Innovation Center of Quantum Information and Quantum Physics, University of Science and Technology of China, Hefei 230026, China}

\author{Min Jiang}
\affiliation{
Hefei National Laboratory for Physical Sciences at the Microscale and Department of Modern Physics, University of Science and Technology of China, Hefei 230026, China}
\affiliation{
CAS Key Laboratory of Microscale Magnetic Resonance, University of Science and Technology of China, Hefei 230026, China}
\affiliation{
Synergetic Innovation Center of Quantum Information and Quantum Physics, University of Science and Technology of China, Hefei 230026, China}

\author{Xin-You L\"u}
\email{xinyoulu@hust.edu.cn}
\affiliation{School of physics, Huazhong University of Science and Technology, Wuhan 430074, China}

\author{Xinhua Peng}
\email{xhpeng@ustc.edu.cn}
\affiliation{
Hefei National Laboratory for Physical Sciences at the Microscale and Department of Modern Physics, University of Science and Technology of China, Hefei 230026, China}
\affiliation{
CAS Key Laboratory of Microscale Magnetic Resonance, University of Science and Technology of China, Hefei 230026, China}
\affiliation{
Synergetic Innovation Center of Quantum Information and Quantum Physics, University of Science and Technology of China, Hefei 230026, China}

\author{Jiangfeng Du}
\affiliation{
Hefei National Laboratory for Physical Sciences at the Microscale and Department of Modern Physics, University of Science and Technology of China, Hefei 230026, China}
\affiliation{
CAS Key Laboratory of Microscale Magnetic Resonance, University of Science and Technology of China, Hefei 230026, China}
\affiliation{
Synergetic Innovation Center of Quantum Information and Quantum Physics, University of Science and Technology of China, Hefei 230026, China}

\begin{abstract}
Superradiant phase transition (SPT) in thermal equilibrium, as a fundamental concept bridging the statistical physics and electrodynamics, can offer the key resources for quantum information science. Notwithstanding its fundamental and practical significances, equilibrium SPT has never been observed in experiments since the first proposal in the 1970s. Furthermore, the existence of equilibrium SPT in the cavity quantum electrodynamics (QED) systems is still subject of ongoing debates, due to the no-go theorem induced by the so-called $A^2$ term.  Based on the platform of nuclear magnetic resonance (NMR), here we experimentally demonstrate the occurrence of equilibrium SPT beyond no-go theorem by introducing the antisqueezing effect. The mechanism relies on the antisqueezing that recovers the singularity of the ground state via exponentially enhancing the zero point fluctuation (ZPF) of system. The strong entanglement and squeezed Schr\"{o}dinger cat states of spins are achieved experimentally in the superradiant phase, which may play an important role in fundamental tests of quantum theory, implementing quantum metrology and high-efficient quantum information processing. Our experiment also shows the antisqueezing-enhanced signal-to-noise rate (SNR) of NMR spectrum, providing a general method for implementing high-precision NMR experiments.
\end{abstract}
\maketitle

Superradiant phase transition, driven by the singularity of quantum fluctuation at the critical point, has undergone tremendous developments in recent years~\cite{Kirton2019}. It was proposed in the Dicke model, describing the collective interaction between $N$ two-level systems and a quantum field, in the thermodynamics limit $N\rightarrow\infty$~\cite{Hepp1973,Wang1973}. When $N$ = 1, the Dicke model is reduced to a Rabi model, in which the SPT has also been predicted theoretically by replacing the thermodynamics limit with the classical oscillator limit $\Omega/\omega\rightarrow\infty$ ($\Omega$ and $\omega$ being the frequency of spin and field, respectively)~\cite{Hwang2015}. Above the quantum critical point, the vacuum (ground state) of cavity field is macroscopically occupied, and becomes twofold degenerate, corresponding to a spontaneously $\mathbb{Z}_2$ symmetry breaking. This leads to the appearance of important quantum effects in the supperradiant phase, including spin-field entanglement, distinguishable quantum superposition with large-amplitude, and so on~\cite{Lambert2004}. 

Cavity (including circuit) QED systems~\cite{Raimond2001,You2011}, allowing to manipulate the light-matter interaction at the quantum level, offer an important platform of implementing SPT. However the required critical parameter regime and ultralow-temperature ground state preparation are normally hard to be satisfied with current technologies of cavity QED. More importantly, the existence of the equilibrium SPT in the cavity QED systems is still challenged by the no-go theorem~\cite{Rzazewski1975,Knight1978,Nataf2010,Viehmann2011,Vukics2014,Jaako2016}.  Specifically, for describing the dipole atom-field interactions in the cavity QED system, the standard Dicke and Rabi Hamiltonians have neglected the squared term of electromagnetic vector potential (i.e., $A^2$ term), which will forbid the occurrence of equilibrium SPT. This is because the $A^2$ term, via adding a coupling-dependent potential of the cavity field, makes the disappearance of the singularity of quantum fluctuation during the whole parameter space.  Until now, the SPT has not been realized experimentally in thermal equilibrium, while the nonequilibrium SPT has been observed in the simulations of the DM with a Bose-Einstein condensate in an open cavity~\cite{Baumann2010,Baden2014,Klinder2015} or a trapped ion setup~\cite{Naini2018}. 
\begin{figure*}
     \centering
     \includegraphics[width=16cm]{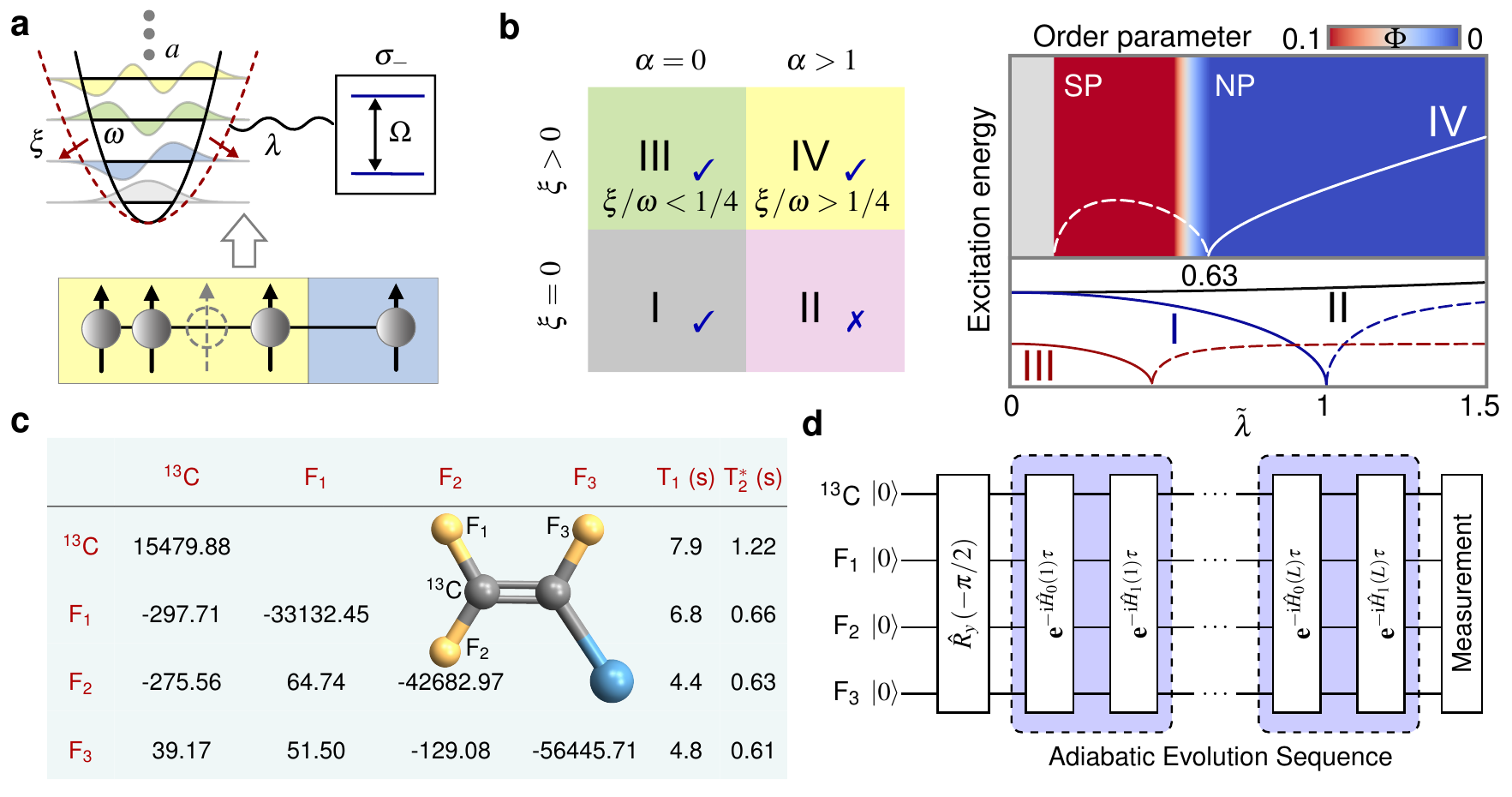}  
     \caption{{\bf Theoretical model, phase diagram, physical system, and quantum circuit for implementing SPT.} {\bf a}, $N$+1 spins are used to map the quantum Rabi model consisting of a two-level system coupled to a Boson field with $2^N$ harmonic levels with coupling strength $\lambda$. The red arrows and dashed potential indicate the antisqueezing effects. {\bf b}, Phase transition property in the limit $\Omega/\omega\rightarrow\infty$, described by the lowest excitation energy and the order parameter $\Phi$ in the parameter regimes from I to IV defined by the values of $\xi$ and $\alpha$. The occurrence and disappearance of SPT are indicated by tick and cross, respectively. The solid and dashed lines correspond to the NP and SP, respectively. {\bf c}, Molecular structure of $^{13}$C-iodotriuroethylene and the relevant system parameters. The diagonal and off-diagonal elements represent chemical shifts and J-couplings, respectively. {\bf d}, Quantum circuit for the adiabatic ground state preparation and the measurement of $\Phi$ with $\hat{H}_0(l)=[1-s(l)]\hat{\mathds{H}}_0$, $\hat{H}_1(l)=s(l)\hat{\mathds{H}}_s$, where $l=0,1,2\dots L$, and $s(l)$ slowly changes from 0 to 1.}
     \label{fig1}
\end{figure*}

Here we employ the quantum simulation technology~\cite{Lloyd1996,Buluta2009,Georgescu2014} to experimentally demonstrate the realization of equilibrium SPT beyond no-go theorem induced by an added antisqueezing of field. We use liquid-state NMR molecules~\cite{Jones2009} to simulate the quantum Rabi model including the $A^2$ and antisqueezing terms (approaching the classical oscillator limit $\Omega/\omega\rightarrow\infty$) by a well-defined spin-to-oscillator mapping scheme. Note that the unaccessible parameter conditions for SPT in the actual cavity QED system can be attained in the platform of NMR. Based on the excellent controllability, the NMR system has been successfully used to simulate topological orders~\cite{luo2018experimentally} and Lee-Yang zeros~\cite{peng2015experimental}. 

Interestingly, we experimentally show that the antisqueezing effect not only makes the appearance of SPT in the case of including $A^2$ term, but also makes the SPT to be reversed, i.e., transition from normal phase (NP) to superradiant phase (SP) along with decreasing spin-field coupling strength. This originally comes from the exponentially enhanced ZPF induced by the antisqueezing effect, that recovers the singularity of the ground state of system. The optimized parameter condition including the necessary antisqueezing strength for phase transition is identified by presenting experimentally the antisqueezing-dependent phase diagram of ground state. In the SP, we experimentally obtain the strong spin-oscillator entanglement and the squeezed Schr\"{o}dinger cat states of spins exhibiting a negative Wigner distribution, large-amplitude separation of peaks, and distinct interference fringes. These states could be used to fault-tolerant quantum computation~\cite{Lund2008,Li2017} and quantum metrology~\cite{Joo2011} approaching Heisenberg limit with actual NMR systems, aside from providing fundamental insights into the nature of decoherence and the quantum–classical transition~\cite{Jeong2003}. Moreover, we show good theory-experiment agreement  due to the antisqueezing-enhanced SNR of NMR spectrum, which applies to the general NMR experiments. Our work also provides the important family of antisqueezing with a new type of applications, besides its widely applications in quantum precision measurement~\cite{Giovannetti2009} and enhancing light-matter interaction~\cite{xinyou2015, Zeytinoglu2017, Leroux2018, Qin2018, Ge2019, Li2020, Chen2020, Burd2020}. 

The Rabi model with Hamiltonian
\begin{align}
\hat{H}_{\rm R}=&\frac{\Omega}{2}\hat{\sigma}_{z}+\omega \hat{a}^{\dagger}\hat{a}+\lambda(\hat{a}^{\dagger}+\hat{a})\hat{\sigma}_{x}
\end{align}
describes a two-level system with frequency $\Omega$ interacting with an oscillator mode with frequency $\omega$, and $\lambda$ denotes the coupling strength. 
Here $\hat{a}$ ($\hat{a}^{\dagger}$) is the annihilation (creation) operator of the oscillator mode, and $\hat{\sigma}_{z}$, $\hat{\sigma}_{x}$ are the Pauli operators for the two-level system. This model has the $\mathbb{Z}_2$ (or parity) symmetry associating with a well-defined parity operator $\hat{\Pi}=e^{i\pi\hat{\mathbb{N}}}$, where $\hat{\mathbb{N}}=\hat{a}^{\dagger}\hat{a}+(1/2)(\hat{\sigma}_z+1)$ is the total excitation number of the system. As shown in the regime ${\rm I}$ of FIG.\,\ref{fig1}{\bf b}, the ground-state SPT is predicted theoretically in the classical oscillator limit $\Omega/\omega\rightarrow\infty$, characterized by a vanishing of the lowest excitation energy~\cite{Hwang2015}. However, this SPT will disappear when the $A^2$ term $\hat{H}_{A}=(\alpha\lambda^2/\Omega)(\hat{a}+\hat{a}^{\dagger})^2$ ($\alpha\geq1$ decided by the Thomas-Reiche-Kuhn sum rule) is included in the actual cavity QED systems, corresponding to the regime ${\rm II}$ of FIG.\,\ref{fig1}{\bf b}. This is the no-go theorem of SPT~\cite{Nataf2010}, and the corresponding debate continues to today from 1970s. Here we will demonstrate experimentally the above no-go theorem can be broken through by introducing an antisqueezing effect, i.e., $\hat{H}_{\rm As}=-\xi(\hat{a}+\hat{a}^{\dagger})^2$ in the platform of NMR. The regime ${\rm IV}$ of FIG.\,\ref{fig1}{\bf b} theoretically shows the reappearance of SPT via the singularity of the excitation energy and the sudden change of the order parameter $\Phi=(\omega/\Omega)\langle \hat{a}^{\dagger}\hat{a}\rangle$ at the critical point $\tilde{\lambda}=\sqrt{1+\alpha\tilde{\lambda}^2-4\xi/\omega}$ with $\tilde{\lambda}=2\lambda/\sqrt{\Omega\omega}$. Specifically, the rescaled ground-state occupation of oscillator $\Phi=(1/4)(\tilde{\lambda}_s^2-\tilde{\lambda}_s^{-2})$ becomes non-zero from $\Phi=0$ at the critical point~\cite{Sup}. The regime ${\rm III}$ of FIG.\,\ref{fig1}{\bf b} demonstrates that the antisqueezing effects could dramatically reduce the critical point of SPT in the case of $\alpha=0$. 

To experimentally demonstrate the ground-state SPT in the platform of NMR, we simulate the Rabi model including the $A^2$ and antisqueezing terms by using $N$+1 spins, as shown in Fig.\,\ref{fig1}{\bf a}. Based on the generators of SU(2), the mapping process is defined as 
 \begin{align}
\hat{a}&=\hat{A}_{-}\sqrt{\hat{\Sigma}_{z}},\,\,\,\,\,\hat{a}^{\dagger}=\hat{A}_{+}\sqrt{ \hat{\Sigma}_{z}+\mathds{1}^{\otimes N}},
\label{eq2} 
  \end{align}
where $\hat{\Sigma}_{z}=-\sum^{N}_{i=1}2^{i-2}\hat{\sigma}^{i}_{z}+(2^N-1)/2$, and $\mathds{1}^{\otimes N}=\mathds{1}\otimes\cdots\otimes\mathds{1} $ (with $N$ spin identity operator $\mathds{1}$) is the identity matrix of $ 2^{N}\times 2^{N} $ dimensions. Here the definitions of operators $\hat{A}_{\pm}$ and the well-defined spin-to-oscillator mapping process are shown in the Methods. This mapping process has some similarities to Holstein-Primakoff transformation, and it is exact in the limit of $N\rightarrow\infty$. We employ \textsuperscript{13}C-iodotriuroethylene dissolved in d-chloroform as 4-qubits system and the experiments are conducted on Bruker Avance III 400 MHz spectrometer at room temperature. The molecule consists of one \textsuperscript{13}C and three \textsuperscript{19}F nuclear spins, as shown in FIG.\,\ref{fig1}{\bf c}. Now the base vectors of 4-qubits span a 16 dimensional Hilbert space $\{|n\rangle\}$ $\left(n=0,1,2....15\right)$, defined in the Methods. In the weak-coupling approximation, the natural Hamiltonian of the sample molecule is described as 
\begin{align}
     \hat{H}_{\rm NMR}=\sum^{4}_{i=1} \pi \omega _{i} \hat{\sigma}_{z}^{(i)} +\sum _{1\leqslant i< j\leqslant 4} \frac{\pi}{2}J_{ij}\hat{\sigma}_{z}^{(i)} \hat{\sigma}_{z}^{(j)},
\end{align}
where $\omega_i$ represents the chemical shift of the $i$-th spin, and $J_{ij}$ is the scalar coupling strength between two spins. The values of parameters $\omega_i$ and $ J_{ij}$ are given in Fig.\,\ref{fig1}{\bf c}. At the beginning, our system, initially at the thermal equilibrium state, is prepared to a pseudo-pure state (PPS) $ \hat{\rho}_{\rm pps} = [(1-\varepsilon)/16]\mathds{1}^{\otimes 4} + \varepsilon\ketbra{0}{0}$ by using the selective-transition approach~\cite{peng2001preparation}. Here $\varepsilon\approx 10^{-5}$ is the polarization, and the fidelity between the experimental initial state and the pure state $|0\rangle \langle 0|$ is 0.991. The detail initialization process is shown in the Supplementary~\cite{Sup}.
\begin{figure}
     \centering
     \includegraphics{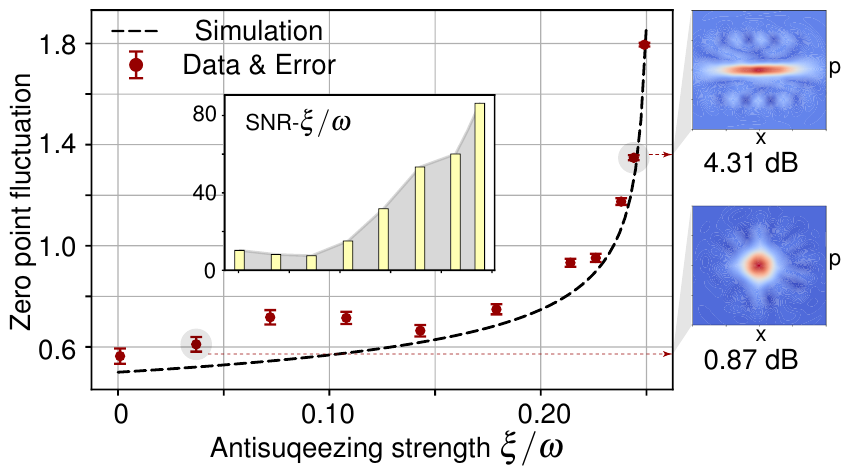}
     \caption{{\bf Experimental demonstration of exponentially enhanced {\rm ZPF} of the oscillator by the antisqueezing effects.} The theoretical (dashed line) and experimental (red circles) {ZPF} versus the antisqueezing strength $\xi$. The inset shows the dependence of the SNR of NMR spectrum on $\xi$. The Wigner functions of two squeezed vacuum states are shown in the right side.}
     \label{fig2}     
\end{figure}

\begin{figure*}
     \centering
     \includegraphics[width=16cm]{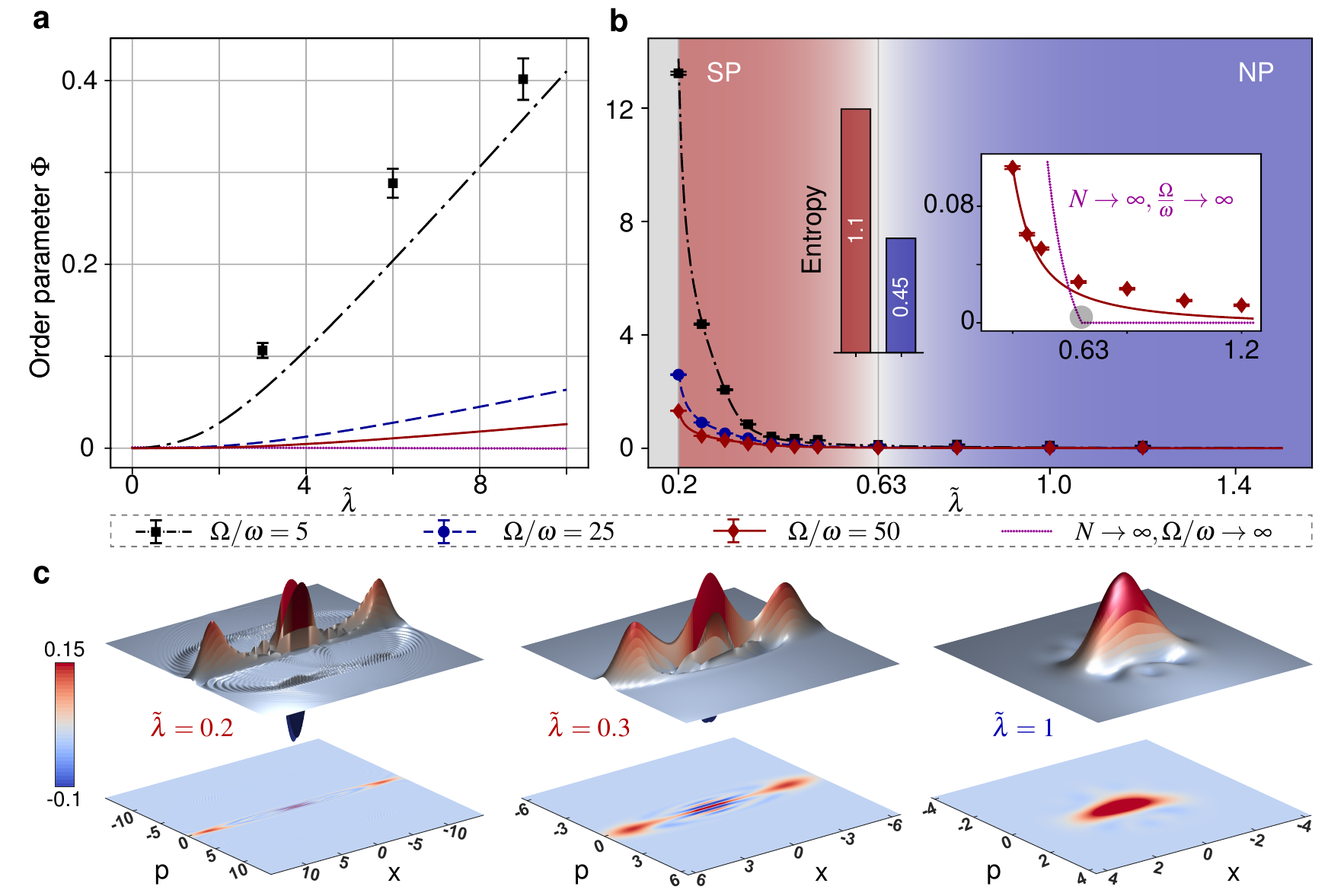}
     \caption{{\bf Experimental demonstration of the recovering of equilibrium SPT induced by the antisqueezing effects.} {\bf{a, b,}} The order parameter $\Phi$ versus the scaled spin-field coupling strength $\tilde{\lambda}$ in the cases of {\bf a} $\xi=0$ and {\bf b} $\xi/\omega=0.26$. The subfigure of {\bf{b}} shows the comparison between the cases of finite-parameter ($\Omega/\omega=50$) and the classical oscillator limit ($\Omega/\omega\rightarrow\infty$), where the SPT occurs exactly. The shading of {\bf{b}} divides NP and SP, according to the exact critical point $\tilde{\lambda}=\sqrt{1+\alpha\tilde{\lambda}^2-4\xi/\omega}$. The bar graph of {\bf b} indicates two von Neumann entropies of system in SP ($\tilde{\lambda}=0.2$) and NP ($\tilde{\lambda}=1$) when $\Omega/\omega=25$. {\bf{c,}} The corresponding Wigner functions of experimental ground states for $\tilde{\lambda}=0.2,\,0.3$ (SP) and $\tilde{\lambda}=1$(NP) when $\Omega/\omega=25$. Here $\alpha=1.1$ is chosen for all subfigures.}
     \label{fig3}     
\end{figure*}

\begin{figure*}
     \centering
     \includegraphics[width=16cm]{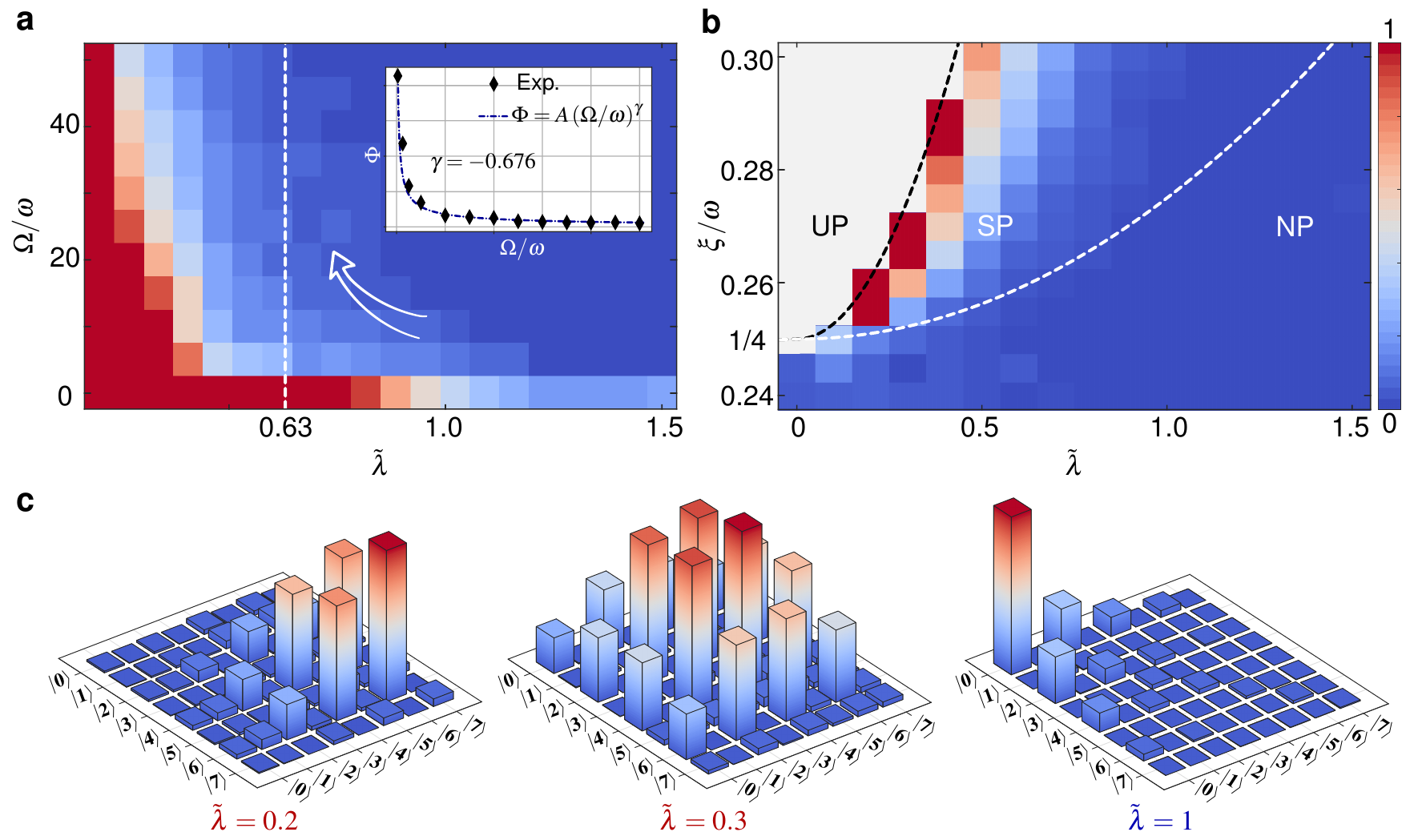}
     \caption{{\bf Antisqueezing modulated ground-state phase diagram.} {\bf{a,}} The order parameter $\Phi$ versus $\tilde{\lambda}$ and $\Omega/\omega$ for $\xi/\omega=0.26$. The inset shows the experimental power-law scaling of $\Phi$ at the critical point, and the fitted finite-frequency scaling exponent is $\gamma=-0.676$. The white dashed line corresponds to the exact critical point in the limit $\Omega/\omega\rightarrow\infty$. {\bf{b,}} The dependence of $\Phi$ on $\tilde{\lambda}$ and $\xi/\omega$ for  $\Omega/\omega=20$. The white (black) dashed line corresponds to the critical parameter distinguishing NP and SP (SP and UP). {\bf{c,}} The reduced density matrix $\hat{\rho}^s_a=\tr_{\sigma}(|G\rangle_s\langle G|)$ for $\Omega/\omega=25$, obtained by the experimentally reconstructed ground state of $\hat{\mathds{H}}_s$. The diagonal elements contribute to $\langle \hat{a}^{\dagger} \hat{a} \rangle_s$, while the sub-sub diagonal elements contribute to $\langle \hat{a}^{\dagger2} \rangle_s$ and $\langle \hat{a}^2 \rangle_s$. The system parameters are same as FIG.\,\ref{fig3}.}
     \label{fig4}     
\end{figure*}

Let us first experimentally demonstrate the antisqueezing-enhanced ZPF in our 4-spins system, which is the key of recovering ground-state SPT in the case of including the $A^2$ term. It is also the nature of antisqueezing enhanced light-matter interaction explored in recent theoretical\,\cite{xinyou2015, Zeytinoglu2017, Leroux2018, Qin2018, Ge2019, Li2020, Chen2020} and experimental\,\cite{Burd2020} works. Theoretically, the antisqueezing term $\hat{H}_{\rm As}$ will make the ground state of an oscillator from a vacuum state $|0\rangle$ to the squeezed vacuum state $\hat{S}(r)|0\rangle$ with $\hat{S}(r)=\exp[r(\hat{a}^2-\hat{a}^{\dagger 2})/2]$ and the squeezing parameter $r=(1/4)\ln(1-4\xi/\omega)$. Then the ZPF, defined as ${\rm ZPF}=\sqrt{\langle \hat{x}^{2} \rangle -\langle \hat{x}\rangle ^{2}}$ with nondimensional quadrature $\hat{x}=(\hat{a}+\hat{a}^{\dagger})/2$, will be exponentially enhanced with increasing the antisqueezing strength $\xi$ (see the dashed line of FIG.\,\ref{fig2}). In our experiment, the truncated squeezing operator in the Hilbert space $\{|n\rangle\}$ is implemented with the gradient ascent pulse engineering (GRAPE) pulses with time 15ms \cite{Khaneja2005}.
The pulse sequences are sequentially applied into the PPS for preparing the squeezing vacuum state. Then, we implement two steps measurement to obtain the ZPF of the squeezing vacuum state from the readout spectra of NMR. (i) Apply four $\pi/2$ readout pulses, i.e., $\hat{R}^{(i)}_y(\pi/2)=\exp(-i\pi/4 \hat{\sigma}^{(i)}_y )$, to reconstruct the diagonal elements \cite{lee2002quantum} of the squeezing vacuum density matrix, corresponding to the expectation $\langle \hat{a}^{\dagger}\hat{a}\rangle$. (ii) Apply identity operator and unitary operator $\hat{U}_1$ (or $\hat{U}_2$) to the fourth (or third) spin, followed by measuring the NMR spectra of the fourth (or third) spin to obtain the expectation $\langle \hat{a} \rangle$ (or $\langle \hat{a}^2\rangle$). Here the operators $\hat{U}_1=\sum^{14}_{n=0}|n\rangle\langle n+1|+|15\rangle\langle 0|$ and $\hat{U}_2=\sum^{13}_{n=0}|{n}\rangle\langle {n+2}|+|14\rangle\langle{0}|+|15\rangle\langle{1}|$ are used to transfer the concerned parts of $\langle \hat{a} \rangle$ and $\langle \hat{a}^2\rangle$ to the single quantum coherence terms, and they can be implemented by the quantum circuits shown in Fig.\,S4 of Supplementary~\cite{Sup} or the equivalent GRAPE pulse sequences. To clearly show the antisqueezing effects, we also present the Wigner functions of two experimentally reconstructed states by the state tomography~\cite{lee2002quantum}. The experimental data shown in FIG.\,\ref{fig2} are in excellent agreement with the theoretical prediction, and the error bars, coming from the statistical fluctuation of the NMR spectra, become smaller along with increasing the antisqueezing effect. This originally comes from the antisqueezing-enhanced SNR of the NMR spectra, as shown in the inset of FIG.\,\ref{fig2} and FIG.\,S5(a) of Supplementary~\cite{Sup}. Our work offers a general method to enhance the measurement precision in the NMR experiments, where the signal is encoded in the $\hat{x}$ quadrature of the detector.   

Next we experimentally show the equilibrium SPT modulated by the antisqueezing effects. The key point is the recovering of ground-state SPT of Rabi model including the $A^2$ term due to the antisqueezing-enhanced ZPF shown above. With the exact squeezing transformation, the ground state of the total system Hamiltonian $\hat{\mathds{H}}=\hat{H}_R + \hat{H}_A +  \hat{H}_{As}$ is equivalent to apply a squeezing operation $\hat{S}(\tilde{r})$ on the ground state of the transformed Hamiltonian 
$\hat{\mathds{H}}_{s}=\hat{S}^{\dagger}(\tilde{r})\hat{\mathds{H}}\hat{S}(\tilde{r})=(\Omega/2)\hat{\sigma}_{z}+\omega_s \hat{a}^{\dagger}\hat{a}+\lambda_s(\hat{a}^{\dagger}+\hat{a})\hat{\sigma}_{x}$~\cite{Sup}. Here $\omega_s=e^{2\tilde{r}}\omega$, $\lambda_s=e^{-\tilde{r}}\lambda$, $\tilde{r}=(1/4)\ln(1+\alpha\tilde{\lambda}^2-4\xi/\omega)$, and the constant term in Hamiltonian $\hat{\mathds{H}}_{s}$ has been ignored. Now the problem is transferred to experimentally preparing the ground state of $\hat{\mathds{H}}_{s}$. In our sample molecule, $^{13}$C spin is labeled as the two-level system, and three $^{19}$F nuclear spins are used to map the truncated boson mode $\hat{a}$ with the defined mapping process Eq.\,(\ref{eq2}). In the experiments, we employ the widely used adiabatic method~\cite{steffen2003experimental} to prepare the ground state of $\hat{\mathds{H}}_s$. According to the quantum circuit shown in FIG.\,\ref{fig1}{\bf d}, the 4-spins sample is firstly prepared into the ground state of Hamiltonian $\hat{\mathds{H}}_0=\sum_{i=1}^4 \hat{\sigma}^{(i)}_y$ by applying $\pi/2$ pulses along $y$ axis on four spins simultaneously. Then the quantum system is controlled to adiabatically evolve under the instantaneous Hamiltonian $\hat{\mathds{H}}(l)=[1-s(l)]\hat{\mathds{H}}_0+s(l)\hat{\mathds{H}}_s $, with $l=1,2,...L$ and $s\left(l\right)$ changing slowly from 0 to 1. The system will finally evolve to the ground state of $\hat{\mathds{H}}_s$, denoted by $|G\rangle_s$, after the above adiabatic evolution sequence. The experimental adiabatic evolution is implemented by the GRAPE pulse with time 26ms. Based on the prepared ground state $|G\rangle_s$, the order parameters of SPT, expressed as $\Phi=(\omega/\Omega)(\cosh(2\tilde{r})\langle \hat{a}^{\dagger} \hat{a} \rangle_s-(1/2)\sinh(2\tilde{r})(\langle \hat{a}^{\dagger2} \rangle_s+ \langle \hat{a}^2 \rangle_s)+ \sinh^2 \tilde{r})$, can be obtained by measuring the corresponding expectations defined with $|G\rangle_s$. The detail measurement processes are shown in the Methods. 

To show the realization of SPT, we present the dependence of the order parameter $\Phi$ (including the theoretical results and experimental data) on $\tilde{\lambda}$ for different frequency ratio $\Omega/\omega$ in FIGs.\,\ref{fig3}{\bf a} and \ref{fig3}{\bf b}. It is shown from FIG.\,\ref{fig3}a that, without antisqueezing effects ($\xi$ = 0), the phase transition is forbidden by the $A^2$ term, i.e., the no-go theorem. However, the SPT is recovered by introducing a fixed antisqueezing effect ($\xi/\omega=0.26$) in FIG.\,\ref{fig3}{\bf b}. Specifically, the experimental order parameter $\Phi$ in FIG.\,\ref{fig3}{\bf b} changes from almost zero to finite number at $\tilde{\lambda}\approx0.63$ with decreasing $\tilde{\lambda}$, which indicates a reversed SPT approximately.  Along with increasing $\Omega/\omega$, this tendency approaches the case of $\Omega/\omega\rightarrow\infty$, where the reversed SPT occurs exactly at the critical point $\tilde{\lambda}=\sqrt{1+\alpha\tilde{\lambda}^2-4\xi/\omega}$ (see the inset of FIG.\,\ref{fig3}{\bf b})~\cite{Sup}. Physically, the occurrence of SPT in our experiment originally comes from the recovering of the singularity of ground state fluctuations due to the antisqueezing-enhanced ZPF of system, as shown in Fig.\,S2 of the Supplementary~\cite{Sup}. Moreover, the comparison between FIGs.\,\ref{fig3}{\bf a} and \ref{fig3}{\bf b} shows that the experimental data agree highly with the theoretical results in the case of introducing the antisqueezing effects, which demonstrates again the property obtained from FIG.\,\ref{fig2}, i.e., the antisqueezing effects can significantly enhance the SNR of the NMR spectra.

Rich quantum resources can be obtained in the superradiant phase, such as the quantum entanglement and quantum superposition of coherent states, i.e., Schr\"{o}dinger cat states. They are significant for quantum metrology and quantum computation, aside from their fundamental nature. For example, Schr\"{o}dinger cat states can enhance the measurement precision by separating the two superposed coherent states with a large distance in phase space, which offers the complementary sensitivity to environmental influences. In the limit $\Omega/\omega\rightarrow\infty$, the ground state of our system (including the antisqueezing term) is theoretically predicted as a squeezed state $|G\rangle_{\rm np}=\hat{S}(\tilde{r}_{\rm np})|0\rangle_a|\!\downarrow\rangle$ in the NP and a spin-oscillator entangled state $|G\rangle_{\rm sp}\approx(1/\sqrt{2})\hat{S}(\tilde{r})[\hat{D}(\beta)|0\rangle_a|\!\downarrow\rangle_{+}+\hat{D}(-\beta)|0\rangle_a|\!\downarrow\rangle_{-}]$ in the SP with a defined displaced operator $\hat{D}(\beta)$~\cite{Sup}. By the state tomography, we experimentally reconstruct the ground states of system, when it is in the NP ($\tilde{\lambda}=1$) and SP ($\tilde{\lambda}=0.2,\,0.3$). The bar graph in FIG.\,\ref{fig3}{\bf b} clearly demonstrate that the strong spin entanglement is obtained in the SP via the von Neumann entropy $S=-\tr(\hat{\rho}_a\log_{2}\hat{\rho}_a)$ ($\hat{\rho}_a$ is the reduced density matrix of oscillator). Moreover, in the SP, the entanglement state $|G\rangle_{\rm sp}$ becomes a squeezed cat state, when we measure the $^{13}$C spin in the $(1/\sqrt{2})(|\!\downarrow\rangle_{+}\pm|\!\downarrow\rangle_{-})$ basis~\cite{Sup}. We plot the corresponding Wigner functions of three experimentally reconstructed ground states in FIG.\,\ref{fig3}{\bf c}, which clearly demonstrate the appearance of squeezed cat states in the SP. They have a negative Wigner distribution with distinct interference fringes and large size for $\tilde{\lambda}=0.3$ and $0.2$, which are the key for implementing super-resolution metrology with high probability and fault-tolerant quantum computing.  Note that, the Schr\"{o}dinger cat state also can be experimentally prepared via the homodyne detection on the number state\cite{Ourjoumtsev2007,Etesse2015}, photon subtraction on the squeezed state\cite{Ourjoumtsev2006,Lo2015}, and high-order nonlinear atom-field interaction\cite{Vlastakis2013}. Here the realization of quantum superposition state indicates a spontaneously $\mathbb{Z}_{2}$ breaking, which is evidenced by the nonzero ground-state coherence $\langle \hat{a}\rangle_{\rm sp}$~\cite{Sup}.  

To fully demonstrate the rich equilibrium dynamics induced by the antisqueezing effect, in FIGs.\,\ref{fig4}{\bf a} and \ref{fig4}{\bf b}, we present the experimental ground-state phase diagram characterized by the rescaled ground-state excitation $\Phi$. The realization of reversed SPT are shown again, and it also can be seen from the reduced density matrix $\hat{\rho}_a^s$ reconstructed experimentally by state tomography. As shown in FIG.\,\ref{fig4}{\bf c}, the main contributions to $\Phi$, i.e., $\langle \hat{a}^{\dagger}\hat{a}\rangle_s$ (diagonal elements) and $\langle \hat{a}^2\rangle_s$ (sub-sub diagonal elements), approximately change from zero to finite value along with decreasing $\tilde{\lambda}$. Figure\,\ref{fig4}{\bf a} again shows that the dependence of $\Phi$ on $\tilde{\lambda}$ approaches to the case of exactly occurring SPT at the quantum critical point along with increasing $\Omega/\omega$. Furthermore, we measure a series of order parameters $\Phi$ at the critical point for different values of $\Omega/\omega$, showing the finite-frequency scaling for the observable $\Phi$ in the inset of FIG.\,\ref{fig4}{\bf a}. The order parameter $\Phi$ vanishes with a power-law scaling, and the fitted finite-frequency scaling exponent $\gamma=-0.676$ is very close to the universal exponent $-2/3$ of the Rabi and Dicke model, which verifies the experimental realization of SPT in finite-frequency regime again. By fixing the value of $\Omega/\omega$, Figure\,\ref{fig4}{\bf b} indicates that $\xi/\omega>1/4$ is required for the occurrence of phase transition, which is consistent with the analytical parameter condition $\tilde{\lambda}\geqslant\sqrt{1+\alpha\tilde{\lambda}^2-4\xi/\omega}$ of SPT. However, too large antisqueezing strength will lead the system enter into the unstable phase (UP), when the rescaled ground-state excitation becomes an imaginary number. With increasing the spin-oscillator coupling $\tilde{\lambda}$, the competition between the $A^2$ and antisqueezing effects will push the system to enter into SP when $\tilde{\lambda}>\sqrt{1+\alpha\tilde{\lambda}^2-4\xi/\omega}$, and then enter into NP when $\tilde{\lambda}<\sqrt{1+\alpha\tilde{\lambda}^2-4\xi/\omega}$.  Our experimental results are approximately agreement with the exact boundaries of different phases, and this consistency becomes better and better along with increasing $\Omega/\omega$ (see FIG.\,S1 of the Supplementary~\cite{Sup}), which again predicts the occurring of exact SPT in the classical oscillator limit. 

In summary, we have presented the first proof-in-principle experimental demonstration of equilibrium SPT beyond no-go theorem induced by the antisqueezing effects. 
To understand the reappearance of SPT, we experimentally shown the enhanced ZPF by antisqueezing, which ultimately recovers the singularity of the ground state of system. The antisqueezing modulated ground-state phase diagram are presented by experimentally preparing the ground state of system with the adiabatic method. Associating with the SPT, we also experimentally realize the strong entanglement and the squeezed cat state of spins, which provides the new possibilities both for quantum metrology and quantum information processing.  Our work is fundamentally interesting in demonstrating that the $A^2$ term is not the ultimate limit for experimental observation of equilibrium SPT in the cavity QED system.

Aside from the above results, the observed antisqueezing-enhanced SNR could be used to the general NMR experiments. The defined spin-to-oscillator mapping process is suitable not only for NMR systems, but will also works well in other spin systems, such as trapped ions\,\cite{Dietrich2003} and NV centers\,\cite{Marcus2013}. They opens the new routes for experimentally exploring the novel quantum optical effects with the platform of NMR or other spin systems.

\clearpage
\section*{Methods}

{\bfseries Spin-to-oscillator mapping scheme.} Generally, $N$ qubits can be used to simulate a boson mode with $2^{N}$ levels by arranging all spin states as the binary form of the corresponding excitation number
\begin{equation}
    \begin{aligned}
        \ket{0}&\mapsto \ket{\uparrow_{N}\uparrow_{N-1}\cdots\uparrow_{1}\uparrow_{0}}=\ket{00\cdots 00},\\
        \ket{1}&\mapsto \ket{\uparrow_{N}\uparrow_{N-1}\cdots\downarrow_{1}\uparrow_{0}}=\ket{00\cdots 01},\\
        \qquad&\vdots\qquad\vdots\\
        \ket{{\rm 2^{N}-1}}&\mapsto \ket{\downarrow_{N}\downarrow_{N-1}\cdots\downarrow_{1}\downarrow_{0}}=\ket{11\cdots 11}.\\
    \end{aligned}
\end{equation}
This scheme makes sure the spin space is fully utilized and the spin matrices are exactly same as the mapped oscillator operators. The mathematical form of this mapping scheme has some similarities to Holstein-Primakoff transformation. We will firstly give the mapping representation of the truncated number operator
\begin{equation}
    \hat{a}^{\dagger}\hat{a}=-\sum^{N}_{i=1}2^{i-2}\hat{\sigma}^{\left(i\right)}_{z}+\frac{2^N-1}{2},\label{number}
\end{equation}
where the superscript $ \left(i\right) $ denotes the $ i $-th qubit. Eq. \eqref{number} can be proved by the mathematical induction.

\emph{Proof of Eq. \eqref{number}}: Obviously, the equation establishes when $ N=1 $. Assume Eq. \eqref{number} is true for $N=k$. Now for $N=k+1$
\begin{equation}
    \begin{aligned}
        &{\rm diag}\left\{0,\cdots,2^{k+1}-1\right\}\\
        &={\rm diag}\left\{0,\cdots,2^{k}-1\right\}\otimes\mathds{1}+2^{k}\mathds{1}^{\otimes k}\otimes\left(\frac{\mathds{1}-\hat{\sigma}_{z}^{\left(k+1\right)}}{2}\right)\\
        &=\left(\frac{2^k-1}{2}-\sum^{k}_{i=1}2^{i-2}\hat{\sigma}^{\left(i\right)}_{z}\right)\otimes\mathds{1}\\
        &\qquad\qquad\qquad\qquad -2^{k-1}\left(\mathds{1}^{\otimes k}\otimes\hat{\sigma}_{z}^{\left(k+1\right)}-\mathds{1}^{\otimes \left(k+1\right)}\right)\\
        &=-\sum^{k+1}_{i=1}2^{i-2}\hat{\sigma}^{\left(i\right)}_{z}+\frac{2^{k+1}-1}{2},
    \end{aligned}
\end{equation}
where $\mathds{1}^{\otimes k}=\mathds{1}\otimes\cdots\otimes\mathds{1}$ is the identity matrix of $2^{k}\times 2^{k}$ dimensions. Then the formula will be true for every natural number $N$.

To obtain the representations of operators $\hat{a}$ and $\hat{a}^{\dagger}$, let's define $\hat{\Sigma}_{z}\equiv\hat{a}^{\dagger}\hat{a}$ with $\hat{\Sigma}_{z}\ket{n}=n\ket{n}$, and the `increasing operator' (`decreasing operator') $ \hat{A}_{+}/$ ($ \hat{A}_{-} $) as follow
\begin{equation}
    \begin{aligned}
        \hat{A}_{+}&\equiv\hat{\sigma}^{\left(1\right)}_{+}+\hat{\sigma}^{\left(2\right)}_{+}\hat{\sigma}^{\left(1\right)}_{-}+\cdots+\hat{\sigma}^{\left(N\right)}_{+}\hat{\sigma}^{\left(N-1\right)}_{-}\cdots\hat{\sigma}^{\left(1\right)}_{-},\\
        \hat{A}_{-}&\equiv\hat{\sigma}^{\left(1\right)}_{-}+\hat{\sigma}^{\left(1\right)}_{+}\hat{\sigma}^{\left(2\right)}_{-}+\cdots+\hat{\sigma}^{\left(1\right)}_{+}\hat{\sigma}^{\left(2\right)}_{+}\cdots\hat{\sigma}^{\left(N\right)}_{-}.
    \end{aligned}
\end{equation}
It is not difficult to find that $ \hat{A}_{+}\ket{n}=\ket{n+1} $ and $ \hat{A}_{-}\ket{n+1}=\ket{n} $ for all $ 0\leq n<2^{N}-1 $. The above definitions allow us to construct truncated creation and annihilation operators conveniently. Based on these properties, we have
\begin{equation}
    \begin{aligned}
        \hat{A}_{-}\sqrt{\hat{\Sigma}_{z}}\ket{n}&=\sqrt{n}\ket{n-1},\\
        \hat{A}_{+}\sqrt{\hat{\Sigma}_{z}+\mathds{1}^{\otimes N}}\ket{n}&=\sqrt{n+1}\ket{n+1}.
    \end{aligned}
\end{equation}
Thus we get the final mapping representations
\begin{equation}
    \begin{aligned}
        \hat{a}&=\hat{A}_{-}\sqrt{\hat{\Sigma}_{z}},\\
        \hat{a}^{\dagger}&=\hat{A}_{+}\sqrt{\hat{\Sigma}_{z}+\mathds{1}^{\otimes N}},
    \end{aligned}\label{truncation}
\end{equation}
which together with Eq. \eqref{number} form the mapping scheme of our quantum simulation process. 

Here we should claim that the form like Eq.\,(\ref{truncation}) is not an easy Hamiltonian to simulate, as there is multi-body interactions. However, by a combination of the natural Hamiltonian of two-body NMR sample and Radio frequency control pulses, multi-body interactions could be simulated in NMR systems~\cite{tseng1999quantum,peng2014experimental}.

{\bfseries Order parameter measurement.} Since $\hat{\mathds{H}}_{s}=\hat{S}^{\dagger}(\tilde{r})\hat{\mathds{H}}\hat{S}(\tilde{r})$, the ground states of $ \hat{\mathds{H}} $ and $ \hat{\mathds{H}}_{s} $ are linked by a squeezing transformation
\begin{equation}
    \ket{G}=\hat{S}\left(\tilde{r}\right)\ket{G}_{s},
\end{equation}
where $ \ket{G} $ and $ \ket{G}_{s} $ are the ground states of $ \hat{\mathds{H}} $ and $ \hat{\mathds{H}}_{s} $, respectively. Then the order parameter of SPT can be expressed as 
\begin{equation}
    \begin{aligned}
        \Phi&=(\omega/\Omega)\langle G|\hat{a}^{\dagger}\hat{a}|G\rangle\\
               &=(\omega/\Omega)\,{_{s}}\langle G|\hat{S}^{\dagger}(\tilde{r})\hat{a}^{\dagger}\hat{a}\hat{S}(\tilde{r})|G\rangle_{s}.
    \end{aligned}
\end{equation}
Together with the following derivation
\begin{equation}
    \begin{aligned}
        \hat{S}^{\dagger}(\tilde{r})&\hat{a}^{\dagger}\hat{a}\hat{S}(\tilde{r})\equiv  \hat{S}^{\dagger}(\tilde{r})\hat{a}^{\dagger}\hat{S}(\tilde{r})\hat{S}^{\dagger}(\tilde{r})\hat{a}\hat{S}(\tilde{r})\\
        &=\cosh(2\tilde{r})\hat{a}^{\dagger}\hat{a}-\frac{1}{2}\sinh(2\tilde{r})\left(\hat{a}^{\dagger 2}+\hat{a}^{2}\right)+\sinh^{2}\tilde{r},
    \end{aligned}
\end{equation}
we obtain 
\begin{align}
\Phi=&(\omega/\Omega)\cosh(2\tilde{r})\langle\hat{a}^{\dagger}\hat{a}\rangle_{s}\nonumber\\
&-(1/2)\sinh(2\tilde{r})\left(\langle \hat{a}^{\dagger 2}\rangle_{s}+\langle \hat{a}^{2}\rangle_{s}\right)+\sinh^{2}\tilde{r}.
\end{align}
This means the order parameters of SPT can be obtained by measuring the corresponding expectations in the ground state $|G\rangle_s$ with the NMR spectra\,\cite{Sup}. 

In short, based on the spin-to-oscillator mapping scheme, i.e., Eq.\,\eqref{number} with $N=3$, the value of $\langle a^{\dagger}a\rangle_s$, corresponding to $\langle\hat{\sigma}_z^{(i)}\rangle_s$, can be obtained by applying three $\pi/2$ readout pulses along $y$ axis on the three \textsuperscript{19}F spins and reading out the corresponding NMR spectra. Similarly, the boson operators $\hat{a}^{\dagger 2}+\hat{a}^{2}$ can be expressed as
\begin{equation}
    \begin{aligned}
        \hat{a}^{\dagger2}+\hat{a}^2=&\sum _{l,m\in \{0,1\}}  \sqrt{ c_{lm} \left(c_{lm}-1\right) }\ketbra{l}{l}\otimes \hat{\sigma}_{x} \otimes \ketbra{m}{m} \\
        &+\sqrt{3\times4}\left( \ketbra{100}{010} + \ketbra{010}{100}\right)\\
        &+\sqrt{4\times5}\left(\ketbra{101}{011} + \ketbra{011}{101}\right).
    \end{aligned}
    \label{Eq-S}
\end{equation}
The expectations of the first term are read out directly from the multiplet structure of F$_2$ spectra. The expectations of the last two terms can be firstly transferred as the single quantum coherence terms of F$_2$ spin by employing operation $\hat{U}_2$ to the system. Subsequently, the corresponding expectations are obtained by measuring the NMR spectra of F$_2$ spin again. Here the operation $\hat{U}_2$ can be implemented by the quantum circuits shown in FIG.\,S4 of Supplementary~\cite{Sup} or the equivalent GRAPE pulse sequences.

\section*{Acknowledgements}
This work is supported by the National Key R \& D Program of China (Grant no. 2016YFA0301203 and No. 2018YFA0306600), the National Science Foundation of China (Grant Nos. 11822502, No. 11974125, No. 11661161018, and No. 11927811), Anhui Initiative in Quantum Information Technologies (Grant No. AHY050000).

\section*{Author contributions}
X. P. and X. L. conceived the project. X. L. conceived the relevant theoretical constructs. X. P., X. C. and Z. W. designed the experiment. X. C. and Z. W. performed the measurements and analyzed the data. M. J. assisted with the experiment. X. P. and J. D. supervised the experiment. All authors contributed to analyzing the data, discussing the results and writing the manuscript. 
\section*{Competing financial interests}
The authors declare no competing financial interests.

\end{document}


\title{Supplementary Material: Experimental quantum simulation of superradiant phase transition beyond no-go theorem via antisqueezing}

\begin{abstract}
This supplementary material contains five parts: I. The detailed derivation of superradiant phase transition (SPT) beyond no-go theorem induced by the antisqueezing effects; II. We analytically present the singularity of quantum fluctuation at the critical point, and the  generation of quantum superposition associating with the SPT. III. The implementation of squeezing operator with the platform of NMR is discussed; IV. We present the detailed procedures of the initial state preparation; V. We show the detail measurement processes for zero point fluctuation (ZPF) and the order parameter; VI. We analyze the experimental errors.
\end{abstract}
\author{Xi Chen}
\thanks{These authors contributed equally to this work}
\affiliation{
Hefei National Laboratory for Physical Sciences at the Microscale and Department of Modern Physics, University of Science and Technology of China, Hefei 230026, China}
\affiliation{
CAS Key Laboratory of Microscale Magnetic Resonance, University of Science and Technology of China, Hefei 230026, China}
\affiliation{
Synergetic Innovation Center of Quantum Information and Quantum Physics, University of Science and Technology of China, Hefei 230026, China}

\author{Ze Wu}
\thanks{These authors contributed equally to this work}
\affiliation{
Hefei National Laboratory for Physical Sciences at the Microscale and Department of Modern Physics, University of Science and Technology of China, Hefei 230026, China}
\affiliation{
CAS Key Laboratory of Microscale Magnetic Resonance, University of Science and Technology of China, Hefei 230026, China}
\affiliation{
Synergetic Innovation Center of Quantum Information and Quantum Physics, University of Science and Technology of China, Hefei 230026, China}

\author{Min Jiang}
\affiliation{
Hefei National Laboratory for Physical Sciences at the Microscale and Department of Modern Physics, University of Science and Technology of China, Hefei 230026, China}
\affiliation{
CAS Key Laboratory of Microscale Magnetic Resonance, University of Science and Technology of China, Hefei 230026, China}
\affiliation{
Synergetic Innovation Center of Quantum Information and Quantum Physics, University of Science and Technology of China, Hefei 230026, China}

\author{Xin-You L\"u}
\email{xinyoulu@hust.edu.cn}
\affiliation{School of physics, Huazhong University of Science and Technology, Wuhan 430074, China}

\author{Xinhua Peng}
\email{xhpeng@ustc.edu.cn}
\affiliation{
Hefei National Laboratory for Physical Sciences at the Microscale and Department of Modern Physics, University of Science and Technology of China, Hefei 230026, China}
\affiliation{
CAS Key Laboratory of Microscale Magnetic Resonance, University of Science and Technology of China, Hefei 230026, China}
\affiliation{
Synergetic Innovation Center of Quantum Information and Quantum Physics, University of Science and Technology of China, Hefei 230026, China}

\author{Jiangfeng Du}
\affiliation{
Hefei National Laboratory for Physical Sciences at the Microscale and Department of Modern Physics, University of Science and Technology of China, Hefei 230026, China}
\affiliation{
CAS Key Laboratory of Microscale Magnetic Resonance, University of Science and Technology of China, Hefei 230026, China}
\affiliation{
Synergetic Innovation Center of Quantum Information and Quantum Physics, University of Science and Technology of China, Hefei 230026, China}

\maketitle

\section{Analytical derivation of antisqueezing induced SPT}
In our work, the simulated cavity QED system including $A^{2}$ and antisqueezing terms is described by Hamiltonian $\hat{\mathds{H}}=\hat{H}_R + \hat{H}_A +  \hat{H}_{As}$, given by
\begin{equation}
    \begin{aligned}
        \hat{H}_R&=\frac{\Omega}{2}\hat{\sigma}_{z}+\omega \hat{a}^{\dagger}\hat{a}+\lambda\left(\hat{a}^{\dagger}+\hat{a}\right)\hat{\sigma}_{x},\\
        \hat{H}_A&=\alpha\frac{\lambda^2}{\Omega}\left(\hat{a}+\hat{a}^{\dagger}\right)^2,\\
        \hat{H}_{As}&=-\xi\left(\hat{a}+\hat{a}^{\dagger}\right)^2,
    \end{aligned}
\end{equation}
where $\hat{H}_R$ is the standard Rabi Hamiltonian, and $\hat{H}_A$ refers to the so-called $A^{2}$-term that prevents the occurrence of equilibrium SPT in normal cavity QED systems. Interestingly, $\hat{H}_{As}$, corresponding to an antisqueezing operation, is the key term to recover SPT in the case of including $ A^{2}$ term. Similar as the case of standard Rabi model~\cite{hwang2015quantum}, Hamiltonian of system $\hat{\mathds{H}}$ can be analytically diagonalized in the classical oscillator limit $\Omega/\omega\rightarrow\infty$, and the corresponding solutions offer the guidance of implementing our experiment shown in the main text.

Applying a squeezing transformation $\hat{S}(\tilde{r})=\exp[\tilde{r}(\hat{a}^2-\hat{a}^{\dagger 2})/2]$ with squeezing parameter
\begin{equation}
   \tilde{r}=\frac{1}{4}\ln\left(1+\alpha\tilde{\lambda}^2-4\frac{\xi}{\omega}\right),\quad \tilde{\lambda}=\frac{2\lambda}{\sqrt{\Omega\omega}},
\end{equation}
into the system Hamiltonian, we obtain
\begin{equation}
    \begin{aligned}
        \hat{\mathds{H}}_s&=\hat{S}^{\dagger}\left(\tilde{r}\right) \hat{\mathds{H}} \hat{S}\left(\tilde{r}\right)\\
        &=\frac{\Omega}{2}\hat{\sigma}_z+\omega_s \hat{a}^{\dagger}\hat{a}+\lambda_s\left(\hat{a}^{\dagger}+\hat{a}\right)\hat{\sigma}_x+C_s,
    \end{aligned}\label{H_s}
\end{equation}
where
\begin{equation}
    \omega_s=\omega e^{2\tilde{r}},\quad\lambda_s=\lambda e^{-\tilde{r}},\quad C_s=\left(e^{2\tilde{r}}-1\right)\frac{\omega}{2}.
\end{equation}
Note that, the ground state of Hamiltonian $\hat{\mathds{H}}$ is equivalent to $\hat{S}\left(\tilde{r}\right)\ket{G}_s$, where $\ket{G}_s$ is the ground state of Hamiltonian $\hat{\mathds{H}}_s$. Now, the effects from $A^2$ term and antisqueezing operation have been introduced into the system parameters $\omega_s$ and $\lambda_s$, and then the property of SPT can be obtained by experimentally searching the ground state of $\mathds{H}_s$, which can be diagonalized in the limit $\Omega/\omega_s\rightarrow\infty$, corresponding to $\Omega/\omega\rightarrow\infty$ in terms of the original system parameter.  

When $\tilde{\lambda}< \exp(2\tilde{r})=\sqrt{1+\alpha\tilde{\lambda}^2-4\xi/\omega}$, the system is in the normal phase (NP). Applying a unitary transformation with
\begin{equation}
    \hat{U}=\exp\left[\frac{\lambda_s}{\Omega} \left(\hat{a}+\hat{a}^{\dagger}  \right)  \left(\hat{\sigma}_{+} -\hat{\sigma}_{-}\right)\right],
\end{equation}
we obtain 
\begin{equation}
    \hat{U}^{\dagger} \hat{\mathds{H}}_s \hat{U} = \omega_s \hat{a}^{\dagger} \hat{a}+\frac{ \tilde{\lambda}_s^2 \omega_s   }{4}(\hat{a}^{\dagger}+\hat{a})^2\hat{\sigma}_z+\frac{\Omega}{2}\hat{\sigma}_z+C_s,
\end{equation}
where $\tilde{\lambda}_s=2\lambda_s/\sqrt{\Omega\omega_s}$. Then, projecting the system into the spin-down subspace, the Hamiltonian becomes
\begin{equation}
    \hat{\mathds{H}}^{\rm np}_s=\omega_s \hat{a}^{\dagger} \hat{a} - \frac{\tilde{\lambda}_s^2 \omega_s}{4}(\hat{a}^{\dagger}+\hat{a})^2-\frac{\Omega}{2}+C_s,
\end{equation}
which can be fully diagonalized through another squeezing transformation
\begin{equation}
    \hat{S}^{\dagger}\left(l_{\rm np}\right) \hat{\mathds{H}}^{\rm np}_s  \hat{S}\left(l_{\rm np}\right) = \omega_e \hat{a}^{\dagger} \hat{a} + E_g.
\end{equation}
Here the squeezing operator $\hat{S}\left(l_{\rm np}\right)=\exp\left[ l_{\rm np} \left( \hat{a}^2 - \hat{a}^{\dagger2}  \right)/2  \right]$
with squeezing parameter $l_{\rm np}= \ln (1-\tilde{\lambda}^2_s)/4$. The excitation energy $\omega_e$ and the ground state energy $E_g$ are given by
\begin{equation}
    \begin{aligned}
        \omega_e &= \omega_s \sqrt{1-\tilde{\lambda}_s^2}  \\
        E_g &= \frac{\omega_s}{2} \left( \sqrt{ 1- \tilde{\lambda}_s^2 } -1  \right) -\frac{\Omega}{2} +C_s.
    \end{aligned}
\end{equation}
The corresponding eigenstates of the system are
\begin{equation}
    \ket{G}_{\rm np}=\hat{S}(\tilde{r}_{\rm np})\ket{m}\ket{\downarrow},\label{Gnp}
\end{equation}
and  $\tilde{r}_{\rm np}=\tilde{r}+l_{\rm np}$. It should be noted that the excitation energy $\omega_e$ is real only for $\tilde{\lambda}< \exp(2\tilde{r})$ and vanishes when $\tilde{\lambda} = \exp(2\tilde{r}) $, which is the typical sign of the occurrence of SPT.
\begin{figure}[H]
    \centering
    \includegraphics{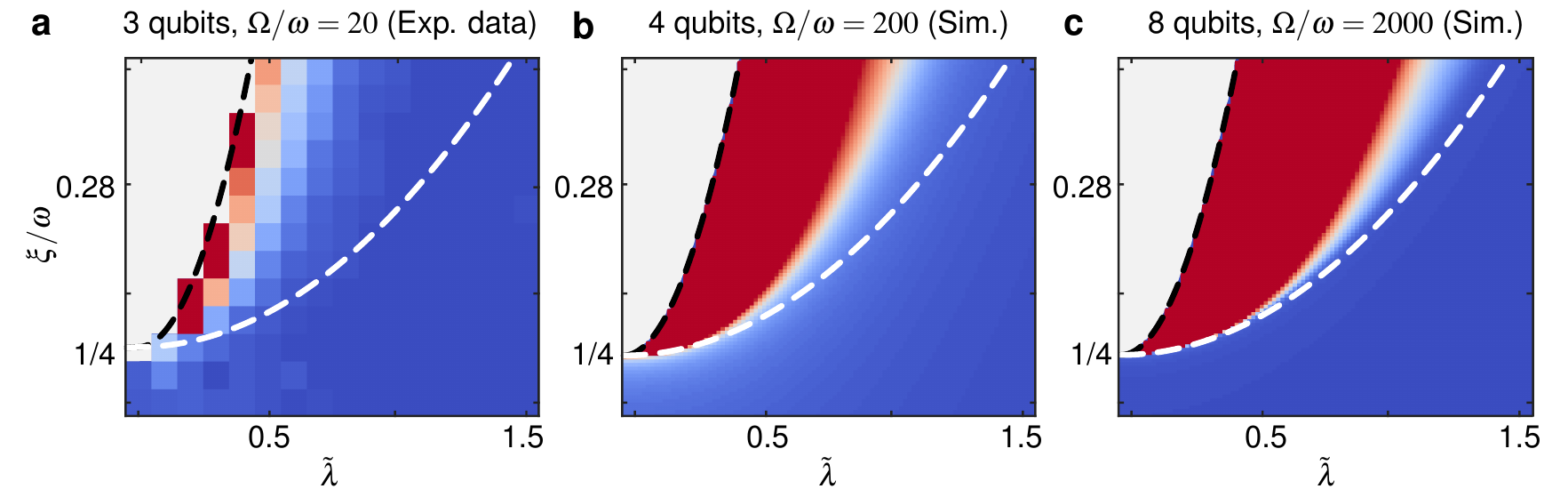}
    \caption{{\bf Asymptotic behavior of the order parameter}. {\bf a,} The dependence of experimental $\Phi$ on $\tilde{\lambda}$ and $\xi/\omega$ for $\Omega/\omega=20$, which is exactly FIG.\,4{\bf b} of main text. {\bf b,} The numerical simulation on 4 qubits and $\Omega/\omega=200$. {\bf c,} The numerical simulation on 8 qubits and $\Omega/\omega=2000$. The results demonstrate that the border of superradiant phase is gradually approaching to the exact critical line (i.e., the white dashed line) when $\Omega/\omega$ and the simulating space becomes larger.}    
    \label{Supp-fig1}
\end{figure}

When $\tilde{\lambda} >\sqrt{1+\alpha\tilde{\lambda}^2-4\xi/\omega}$, the system enters into the superradiant phase (SP), and the Hamiltonian $\hat{\mathds{H}}^{\rm np}_s$ becomes invalid since the boson mode $\hat{a}$ are macroscopically excited. Then we should firstly apply the displacement transformation into Hamiltonian $\hat{\mathds{H}}_s$ and obtain  
\begin{equation}
\hat{D}^{\dagger}(\beta)\hat{\mathds{H}}^{\rm np}_s\hat{D}(\beta)=\frac{\Omega'}{2}\hat{\sigma}'_z+\omega_s \hat{a}^{\dagger}\hat{a}+\lambda_s\left(\hat{a}^{\dagger}+\hat{a}\right)\hat{\sigma}'_x+\omega_s\beta^2+C_s,\label{DH}
\end{equation}
where $\Omega'=\tilde{\lambda}^2_s\Omega$, $\hat{D}\left(\beta\right)=\exp\left(\beta \hat{a}^{\dagger}-\beta^{*}\hat{a}\right)$ with $\beta = \pm\sqrt{\Omega\left( \tilde{\lambda}_s^2 - \tilde{\lambda}_s^{-2} \right)/4 \omega_s}$, and the rotated Pauli operators $\hat{\sigma}'_z$, $\hat{\sigma}'_x$ are given by $|\downarrow\rangle'=\cos\theta|\downarrow\rangle-\sin\theta|\uparrow\rangle$ and $|\uparrow\rangle'=\sin\theta|\downarrow\rangle+\cos\theta|\uparrow\rangle$ with $\tan(2\theta)=-4\lambda_s\beta/\Omega$. Hamiltonian (\ref{DH}) is similar as Hamiltonian $\hat{\mathds{H}}_s$, and then it can be diagonalized by employing the similar procedure used in the normal phase. The corresponding excitation energy $\omega'_e$ and ground state energy $E'_g$ becomes
\begin{equation}
    \begin{aligned}
        \omega'_e &= \omega_s \sqrt{1-\tilde{\lambda}_s^{-4}}  \\
        E'_g &= \frac{\omega_s}{2} \left( \sqrt{ 1- \tilde{\lambda}_s^{-4}} -1  \right) -\frac{\Omega}{4}(\tilde{\lambda}^2_s+\tilde{\lambda}^{-2}_s) +C_s.
    \end{aligned}
\end{equation}
The ground state of system becomes twofold degenerate, i.e.,
\begin{equation}
    \begin{aligned}
    \ket{G}^{\pm}_{\rm sp}&=\hat{S}(\tilde{r})\hat{D}(\pm\left|\beta\right|)\hat{S}(l_{\rm sp})\ket{0}_a\ket{\downarrow}_{\pm}\approx\hat{S}(\tilde{r})\hat{D}(\pm\left|\beta\right|)\ket{0}_a\ket{\downarrow}_{\pm},
    \end{aligned}
\end{equation}
where $l_{\rm sp}=(1/4)\ln\left(1-\tilde{\lambda}^{-4}_s\right)$ approaches zero in the parameter regime chosen in our experiment, and 
\begin{equation}
    \begin{aligned}
        \ket{\downarrow}_{\pm}&=\sqrt{\frac{1}{2}\left(1+\tilde{\lambda}^{-2}_s\right)}\ket{\downarrow}\pm\sqrt{\frac{1}{2}\left(1-\tilde{\lambda}^{-2}_s\right)}\ket{\uparrow}.
    \end{aligned}
\end{equation}
Then the ground state of system in the superradiant phase can be written as
\begin{equation}
    \begin{aligned}
    \ket{G}_{\rm sp}&\approx\frac{1}{\sqrt{2}}\hat{S}(\tilde{r})\left[\hat{D}(\left|\beta\right|)\ket{0}_a\ket{\downarrow}_{+}+\hat{D}(-\left|\beta\right|)\ket{0}_a\ket{\downarrow}_{-}\right].\label{Gsp}
    \end{aligned}
\end{equation}

The SPT also can be characterized by the sudden change of the rescaled ground-state occupation of field $\hat{a}$, i.e., $\Phi=(\omega/\Omega)\langle \hat{a}^{\dagger}\hat{a}\rangle_g$. Based on the above analytical results in the classical oscillator limit $\Omega/\omega\rightarrow\infty$, $\Phi=0$ when the system is in the normal phase $\left(\tilde{\lambda}<\sqrt{1+\alpha\tilde{\lambda}^2-4\xi/\omega}\right)$, and $\Phi=(e^{-4\tilde{r}}/4)(\tilde{\lambda}_s^2-\tilde{\lambda}_s^{-2})$ becomes non-zero when it enters into the superradiant phase $\left(\tilde{\lambda}>\sqrt{1+\alpha\tilde{\lambda}^2-4\xi/\omega}\right)$. Associating with the SPT, there is a spontaneously $\mathbb{Z}_2$ symmetry breaking, i.e., the $\mathbb{Z}_2$ symmetry of the ground state is broken in the superradiant phase. It can be demonstrated by the ground-state coherence of field $\langle \hat{a}\rangle_g$, and $\langle \hat{a}\rangle_{g}$ changes from $0$ to $\pm \exp(-\tilde{r})|\beta|$ when the system enters into the superradiant phase from normal phase. Besides the normal and superradiant phases, the system will become unstable when $\omega_s$ and $\Phi$ becomes the imaginary numbers in the regime of $1+\alpha\tilde{\lambda}^2-4\xi/\omega<0$, corresponding to $\tilde{\lambda}<\sqrt{(1/\alpha)(4\xi/\omega-1)}$. In a short summary, the order parameter $\Phi$ will sudden changes at the critical lines $\tilde{\lambda}=\sqrt{1+\alpha\tilde{\lambda}^2-4\xi/\omega}$ and $\tilde{\lambda}=\sqrt{(1/\alpha)(4\xi/\omega-1)}$, which divides the NP, SP and unstable phase (UP). In the finite parameter regime, the dependence of $\Phi$ on the system parameters approaches the case of exact phase transition along with increasing $\Omega/\omega$, and this asymptotic behavior is clearly shown in FIG.\,\ref{Supp-fig1}.

\section{ZPF singularity and quantum superposition associating with the SPT}
\begin{figure}[H]
    \centering
    \includegraphics{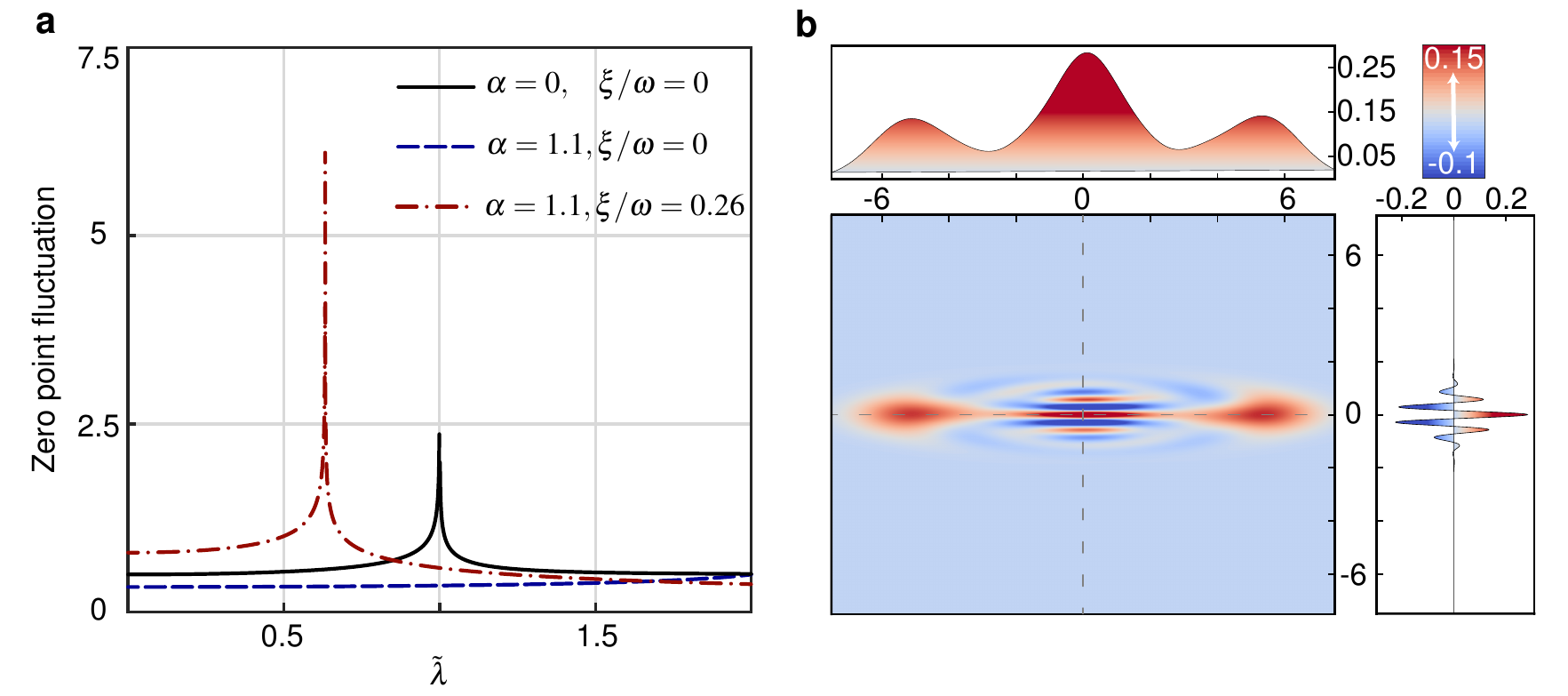}
    \caption{{\bf Antisqueezing recovering the  singularity of ZPF and experimental Schr\"{o}dinger cat states in SP.} {\bf a,} The black line demonstrates the singularity of Rabi model while the blue line means the disappearance of the singularity because of no-go term, and the antisqueezing effect recovers it as shown by the red line. {\bf b,} The Wigner functions of experimental reconstructed ground state with system parameters $\Omega/\omega=25$, $ \tilde{\lambda}=0.3$.}   
    \label{Supp-fig2} 
\end{figure}

The above results show that the antisqueezing effect could induce the occurrence of SPT even in the case of including $A^2$ term. This essentially comes from the recovering of singularity of quantum fluctuation at the critical point induced by the antisqueezing effects. Based on the analytical solutions in Section I, this singularity property could be demonstrated by calculating the ZPF of system. Firstly, the variance of quadrature $\hat{x}=\left( \hat{a} + \hat{a}^{\dagger} \right)/2$ in the ground state of Rabi model with Hamiltonian $\hat{H}_{R}$ is
\begin{equation}
    {\rm ZPF}_{1}\equiv\sqrt{ \expval{\hat{x}^2} - \expval{\hat{x}}^2 }=
    \begin{cases}
        (1/2)\left( 1-\tilde{\lambda}^2  \right)^{-1/4} \qquad &\left(\tilde{\lambda}<1\right),\\
        (1/2)\left( 1-\tilde{\lambda}^{-4}  \right)^{-1/4}\qquad &\left(\tilde{\lambda}>1\right),
    \end{cases}
\end{equation}
which corresponds to the black solid line of FIG.\,\ref{Supp-fig2}(a). Secondly, including the $A^2$ term, the Hamiltonian of system becomes $\hat{H}_{R}+\hat{H}_{A}$, and the ZPF is
\begin{equation}
    {\rm ZPF}_{2}=\frac{1}{2} \left(1+\alpha \tilde{\lambda}^2 -\tilde{\lambda}^2  \right)^{-\frac{1}{4}},
\end{equation}
which is the blue dashed line of FIG.\,\ref{Supp-fig2}(a), and clearly demonstrates that the singularity of ZPF disappears, corresponding to the no-go theorem. Lastly, with the antisqueezing term $H_{As}$ included, the ZPF of system with Hamiltonian $\hat{H}_{R}+\hat{H}_{A}+\hat{H}_{As}$ becomes
\begin{equation}
    {\rm ZPF}_{3}=
    \begin{cases}
        (1/2)\left( 1-\tilde{\lambda}_s^2  \right)^{-1/4}  \left( 1+\alpha\tilde{\lambda}^2-4\xi/\omega \right)^{-1/4}\qquad &\left(\tilde{\lambda}<\tilde{\lambda}_{c}\right),\\
        (1/2)\left( 1-\tilde{\lambda}_s^{-4}  \right)^{-1/4}  \left( 1+\alpha\tilde{\lambda}^2-4\xi/\omega \right)^{-1/4}\qquad &\left(\tilde{\lambda}>\tilde{\lambda}_{c}\right),
    \end{cases}
\end{equation}
which corresponds to the red dot-dash line in FIG.~\ref{Supp-fig2}(a), and clearly shows that the singularity of ZPF is recovered. Here $\tilde{\lambda}_c=\sqrt{ \left( 4 \xi/\omega -1 \right) /\left( \alpha-1 \right) }$, obtained by $\tilde{\lambda}=\sqrt{1+\alpha\tilde{\lambda}^2-4\xi/\omega}$. This results also certify the main methodology of this work that the antisqueezing effect can break through the no-go theorem of SPT. 

Associating with the SPT, quantum entanglement and quantum superposition occur in the superradiant phase. In the limit $\Omega/\omega\rightarrow\infty$, the ground state of Hamiltonian $\mathds{H}$ is theoretically predicted as a spin-field entangled state in the superradiant phase as Eq.~\eqref{Gsp} shows. From this entangled ground state, we also can obtain the Schr\"{o}dinger cat states of field $\hat{a}$ by measuring the spin in the $(|\downarrow\rangle_{+}\pm|\uparrow\rangle_{-})/\sqrt{2}$ basis. Depending on the outcome of the measurement, the state of simulated boson field is approximately projected into one of the following squeezed cat states
\begin{equation}
    \begin{aligned}
    \ket{\Psi}_{\rm cat}&\approx\frac{1}{\sqrt{2}}\hat{S}(\tilde{r})\left[\hat{D}(|\beta|)\ket{0}_a\pm\hat{D}(-|\beta|)\ket{0}_a\right].\label{cat}
    \end{aligned}
\end{equation}
In FIG.\,\ref{Supp-fig2}(b), we plot the Wigner function of the experimental reconstructed ground state, which is a clear squeezed Schr\"{o}dinger cat state as the theoretical prediction. Also the obtained Schr\"{o}dinger cat state has the negative Wigner distribution and distinct interference fringes. Two superposed coherent states separate with a large distance in phase space. This means it offers good quantum resources for quantum precision measurement and quantum information processing.

\section{The implementation of squeezing operator}
The antisqueezing effect plays a pivotal role in recovering SPT with the presence of $A^2$ term. Based on the defined mapping scheme in the Methods of the main text, the truncated squeezing matrices in the $2^{N}$ dimensional Hilbert space can be mapped by using $N$ spins, i.e., $\hat{S}(r)=\exp[r( \hat{a}^2-\hat{a}^{\dagger2})/2]$, where $\hat{a}$ and $\hat{a}^{\dagger}$ are given by  Eq.\,(3) of main text. Take $N = 4$ for example, We have
\begin{equation}
    \begin{aligned}
        \hat{a}^2-\hat{a}^{\dagger2} &= \sum_{n=0}^{2^N-1} \sqrt{(n+1)  (n+2) } \ketbra{ n}{n+2}-h.c.
    \end{aligned}
    \label{Eq-S}
\end{equation}
Then the truncated squeezing matrices can be implemented by NMR pulse sequences with time 15 ms, which are optimized by gradient ascent pulse engineering (GRAPE) method.

\begin{figure}[H]
    \centering
    \includegraphics[width=0.92\textwidth]{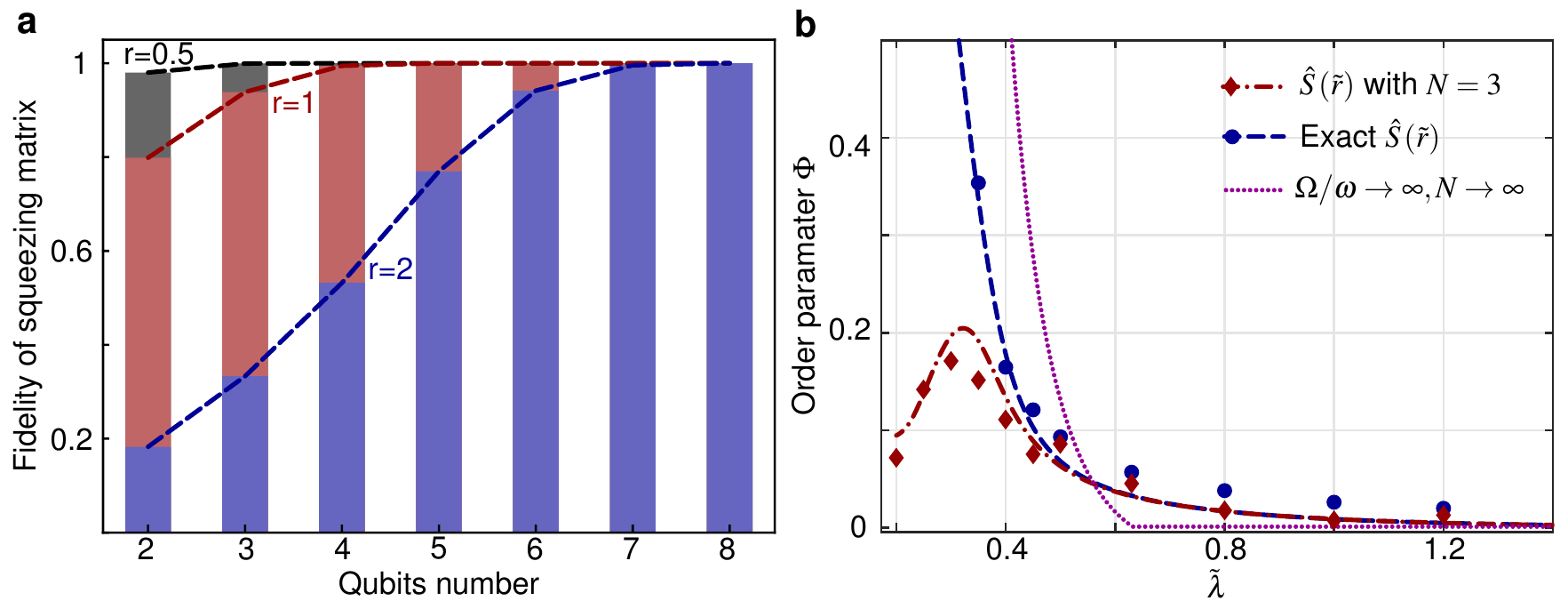}
    \caption{{\bf Validity of the truncated squeezing operator.} {\bf a,} The fidelities of the squeezing operator vs the used qubit number for different squeezing parameters $r$. Here 10-qubit squeezing operator is chosen as the target matrix. {\bf b,} The dependences of order parameter $\Phi$ on $\tilde{\lambda}$ for the cases of $\hat{S}(\tilde{r})$ with $N=3$, the exact $\hat{S}(\tilde{r})$, and the theoretical prediction in the classical oscillator limit. Here the diamonds and circles are the experimental data, and the system parameters are $\Omega/\omega=25$ and $\xi/\omega=0.26$.} \label{Supp-fig3}   
\end{figure}

To identify quantitatively the valid regime of the mapped squeezing operator, we define the fidelity of the truncated squeezing operator as $F=\langle\psi|\hat{S}^{\dagger}_{T}(r)\hat{S}(r)|\psi\rangle$, where $\hat{S}_{T}(r)$ and $\hat{S}(r)$ are the target matrix and the mapped matrix with $N$ spins, respectively. In principle, $|\psi\rangle$ is any pure state of boson field $\hat{a}$, and here we choose the vacuum state for simplification. In FIG.\,\ref{Supp-fig3}, we present the fidelities of the mapped squeezing operators with different squeezing parameters. Here the target matrix is set to the squeezing operator truncated in $2^{10}\times2^{10}$ dimensional Hilbert space since it has been very close to the squeezing operator in the finite dimensional Hilbert space. It is shown from FIG.\,\ref{Supp-fig3}(a) that the fidelities become very low in the case of small $N$ along with increasing the squeezing parameter $r$. To avoid this problem and accurately demonstrate the SPT property with our 4-spins sample, we experimentally prepare the ground state $|G\rangle_s$ of the transformed Hamiltonian $\hat{\mathds{H}}_s=\hat{S}^{\dagger}(\tilde{r})\hat{\mathds{H}}\hat{S}(\tilde{r})$. Then the ground state of system can be obtained by $|G\rangle=\hat{S}({\tilde{r}})|G\rangle_s$ with $\hat{S}({\tilde{r}})$ being the exact squeezing operator. This method has been used to demonstrate the SPT in the main text, and it effectively utilizes the resource of NMR sample with finite spins. Moreover, FIG.\,\ref{Supp-fig3}(b) also shows that there does not exist qualitative difference between the cases of mapped squeezing operator with $3$ spins sample and the exact squeezing operator, and both of them are approximately consistent with the theoretical prediction in the classical oscillator limit $\Omega/\omega\rightarrow\infty$ and $N\rightarrow\infty$, where the SPT occurs exactly.         

\section{Initial state preparation}
At the beginning, the NMR system is at thermal equilibrium state
\begin{equation}
     \hat{\rho}_{\rm eq}=\frac{\mathds{1}^{\otimes 4}}{2^4} +  \frac{\hbar  B_0}{2^4 k_B T} \sum ^{4}_{i=1}  \gamma_i \hat{\sigma}_z^{\left(i\right)}
\end{equation}
where $\mathds{1}^{\otimes 4}$ represents the $2^4 \times 2^4$ identity operator, $B_0=9.6$ Tesla is the static magnetic field, $T$ is the room temperature 300K, and $\gamma_i$ is the Gyromagnetic ratio of $i$-th spin. Note that this thermal equilibrium state is a highly mixed state due to $(\hbar B_0  \gamma_i)/(2^4 k_B T)\approx10^{ - 5}$. At the beginning of our experiment, we need to initialize the 4-spins system as the pseudo-pure state (PPS),
\begin{equation}
     {\hat{\rho}_{\text{pps}}} = \frac{{1 - \varepsilon }}{{16}}{\mathds{1}^{\otimes 4}} + \varepsilon \left| {0000} \right\rangle \left\langle {0000} \right|  
     \label{Eq.pps}
\end{equation}
where $\varepsilon  \approx {10^{ - 5}}$ is the polarization.
Follow the standard convention in ensemble quantum computing~\cite{gershenfeld1997bulk} , we neglect the identity part in the PPS since it has no influence on NMR signals. From now on, we regard PPS as a true vacuum state.

To prepare the PPS, we use selective-transition approach~\cite{peng2001preparation}, where some unitary matrices and two field gradient pulses along $z$ direction (Gz) are implemented. Field gradient pulses are used to implement non-unitary operation. Notice that the PPS has a diagonal density matrix, where the first diagonal element has a relative large value and the others are the same. So we should redistribute the diagonal elements from equilibrium state while keep the off-diagonal elements as zeros. The first step is applying 14 single-transition unitary operators $\hat{U}=\exp(-i\beta_{jk}\hat{u}_{jk})$ to equilibrium state to redistribute the populations, where $j$ and $k$ represent different energy levels and $\hat{u}_{jk}=( |j\rangle \langle k| + |k\rangle \langle j|)/2$. The transition angle $\beta_{jk}$ is optimized by numerical search procedure. After these operations, the diagonal elements are well distributed, but many undesired off-diagonal elements (coherence terms) come out. Then a following Gz pulse is used to eliminate the coherence terms except the zero-quantum coherences of three homonuclear \textsuperscript{19}F spins. Next, these zero coherence terms are transferred to other no-zero coherence terms, followed by the second Gz pulse. Now the PPS is prepared and the full tomography~\cite{lee2002quantum} results show the fidelity between the experimental PPS and the pure state $|0000\rangle \langle 0000|$ is over 0.99. Here the fidelity between states $\hat{\rho}_1$ and $\hat{\rho}_2$ is defined as
\begin{equation}
    F( \hat{\rho}_{1},\hat{\rho}_{2}) =\frac{\tr(\hat{\rho}_{1} \hat{\rho}_{2})}{\sqrt{\tr\left( \hat{\rho}^{2}_{1}\right) \tr\left( \hat{\rho}^{2}_{2}\right)}}.
\end{equation}

\begin{figure}[H]
    \centering
    \includegraphics[width=0.55\textwidth]{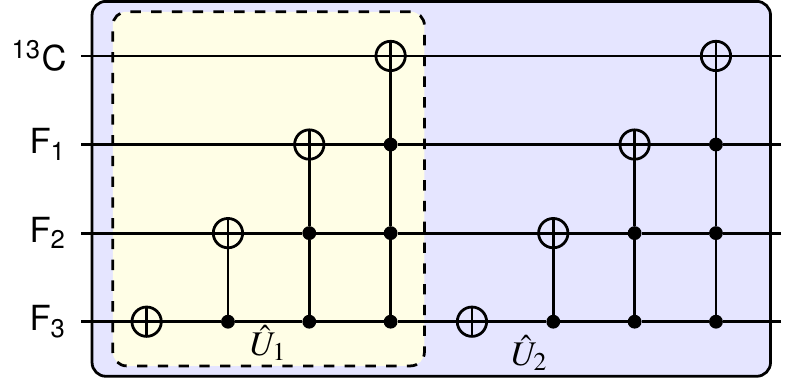}
    \caption{{\bf Measurement circuits.} The quantum circuit for operations $\hat{U}_1$ (dashed box) and $\hat{U}_2$ ($\hat{U}_2=\hat{U}_1^2$). }    
    \label{fig:Supp-fig4}
\end{figure}

\section{Measurement of the ZPF and order parameter}
The ZPF of simulated field and the order parameters of SPT are measured in our experiment. ZPF is defined as ${\rm ZPF}=\sqrt{\langle \hat{x}^{2} \rangle -\langle \hat{x}\rangle ^{2}}$ with $\hat{x}=\left(\hat{a}+\hat{a}^{\dagger}\right)/2 $. In our experiments, we use four spins to simulate $2^4=16$ dimension truncated boson space, and measure $\tr(\hat{\rho} \hat{x}^2) $, $\tr(\hat{\rho}\hat{x}) $ respectively. According to our mapping proposal, $\hat{x}$ is described in the mapped basis as 
\begin{equation}
    \hat{a}+\hat{a}^{\dagger}=\begin{pmatrix}
        & \bra{0} & \bra{1} & \bra{2} & \bra{3} & \cdots & \bra{15} \\
        \ket{0} & 0 & 1 & 0 & 0 & \cdots & 0 \\
        \ket{1} & 1 & 0 & \sqrt{2}& 0 & \cdots & 0 \\
        \ket{2} & 0 & \sqrt{2} & 0 & \sqrt{3} & \cdots & 0 \\
        \ket{3} & 0 & 0 & \sqrt{3} & 0 & \cdots & 0 \\
        \vdots & \vdots & \vdots & \vdots & \vdots & \cdots & \sqrt{15} \\
        \ket{15} & 0 & 0 & 0  & \cdots& \sqrt{15} & 0 
    \end{pmatrix},
\end{equation}
which means
\begin{equation}
    \begin{aligned}
        \hat{a}+\hat{a}^{\dagger} &= \sum_{n=0}^{15} \sqrt{n+1}| n+1\rangle \langle n|+ h.c.\\
      &=\sqrt{1} | 1 \rangle \langle 0| + \sqrt{2} | 2 \rangle \langle 1| +
      \sqrt{3}| 3 \rangle \langle 2|+\cdots + \sqrt{15}| 15 \rangle \langle 14| + h.c.
    \end{aligned} 
    \label{Eq. adagger_pluse_a}
\end{equation}
The first, third, fifth \dots terms in Eq.\,(\ref{Eq. adagger_pluse_a}) are the single coherence terms, and the corresponding expectations can be read out directly from the NMR spectra of the fourth spin. The second, fourth, sixth\dots terms are firstly transferred to the single coherence terms through the operation $\hat{U}_1=\sum^{14}_{n=0}|{n}\rangle\langle{n+1}|+|{15}\rangle\langle{0}|$, and then they are obtained by reading the NMR spectra of the fourth spin. An effective quantum circuit to implement $\hat{U}_1$ is shown in FIG.\,\ref{fig:Supp-fig4}, and a linear combination of the above obtained expectations gives the value of $\tr(\hat{\rho}\hat{x})$. The scheme to obtain $\tr(\hat{\rho}\hat{x}^2) $ is similar. Note that $\hat{x}^2= ({\hat{a}^{\dagger2}} +\hat{a}^2 + \hat{a}^{\dagger} \hat{a} + \hat{a}\hat{a}^{\dagger})/4$, and we have 
\begin{equation}
    \hat{a}^2+\hat{a}^{\dagger2}=\begin{pmatrix}
        & \bra{0} & \bra{1} & \bra{2} & \bra{3} & \cdots & \bra{15} \\
        \ket{0} & 0 & 0 & \sqrt{1\times2} & 0 & \cdots & 0 \\
        \ket{1} & 0 & 0 & 0 & \sqrt{2\times 3} & \cdots & 0 \\
        \ket{2} & \sqrt{1\times2} & 0 & 0 & 0 & \cdots & 0 \\
        \ket{3} & 0 & \sqrt{2\times3} & 0 & 0 & \cdots & \sqrt{14\times 15} \\
        \vdots & \vdots & \vdots & \vdots &\vdots  & \cdots  & 0\\
        \ket{15} & 0 & 0   & \cdots& \sqrt{14\times 15}& 0 & 0 
    \end{pmatrix},
\end{equation}
which means
\begin{equation}
    \begin{aligned}
        \hat{a}^2+\hat{a}^{\dagger2} &= \sum_{n=0}^{13} \sqrt{(n+1)   \times (n+2) }| n+2 \rangle \langle n|+h.c.\\
      &=\sqrt{1\times 2} | 2 \rangle \langle 0| + \sqrt{2\times 3} | 3 \rangle \langle 1| +
      \sqrt{3 \times 4}| 4 \rangle \langle 2|+\cdots + \sqrt{14\times 15}| 15 \rangle \langle 13| + h.c.
    \end{aligned} 
    \label{Eq. adagger2_pluse_a2}
\end{equation}
Similar to the case of measuring $\langle \hat{a} + \hat{a}^{\dagger}\rangle$, a half of terms in Eq.\,(\ref{Eq. adagger2_pluse_a2}) are the single coherence for the third spin, and can be read directly from its NMR spectra. For reading out the left expectations of operators, a readout operator $\hat{U}_2$ is implemented before measuring the third spin. Here $\hat{U}_2=\sum^{13}_{n=0}|{n}\rangle\langle{n+2}|+|{14}\rangle\langle{0}|+|{15}\rangle\langle{1}|$ and it transfers the left terms of Eq.\,(\ref{Eq. adagger2_pluse_a2}) to the single coherence terms. Combining the observed expectations with the coefficients in Eq.\,(\ref{Eq. adagger2_pluse_a2}), the value of $ \expval{ \hat{a}^2+\hat{a}^{\dagger2}} $ is obtained. Moreover, another four readout pulses $\hat{R}^{(i)}_{y}(\pi/2)=\exp(-i\pi/4 \hat{\sigma}^{(i)}_{y})$ are needed to reconstruct the diagonal elements to obtain $\expval{\hat{a}^{\dagger}\hat{a}} $. Combining all the measured expectations gives the $\tr(\hat{\rho}\hat{x}^2) $.

Next, let us describe the experimental details of measuring the order parameter of SPT. As mentioned in the Methods, the order parameters of SPT can be obtained by measuring $\expval{\hat{a}^{\dagger}\hat{a}}_s$ and $\expval{\hat{a}^{\dagger 2}}_s + \expval{\hat{a}^{2}}_s$ with ground state $\ket*{ G}_s$. To get the result of $\expval{\hat{a}^{\dagger}\hat{a}}_s $, we apply three $\pi/2$ readout pulses along $y$ axis to three \textsuperscript{19}F spins, then read out the NMR spectra to obtain the corresponding expectations $\expval{\hat{\sigma}^{(i)}_z}$, respectively. The corresponding NMR spectra are the lines $1$-$3$ of FIG.\,\ref{Supp-fig5}(b). For measuring $\expval{\hat{a}^{\dagger 2}}_s +  \expval{\hat{a}^{2}}_s$, we need two readout spectra. The first one is obtained by reading out the third spin directly, and the spectra is the line 5 of FIG.\,\ref{Supp-fig5}(b). The other one is obtained by applying operation $\hat{U}_2$ to the system, and then reading out the third spin. The corresponding spectra is the line 4 of FIG.\,\ref{Supp-fig5}(b).
\begin{table}[H]
    \setlength\arrayrulewidth{0.8pt}
    \setlength\tabcolsep{8pt}
    \centering
    \begin{tabular}{|c|c|c|c|c|}
    \hline
    \hline
       $ \xi $ & $ \mathds{D}_{\rm tot} $ & $ \mathds{D}_{\rm pps} $ & $ \mathds{D}_{\rm deco} $ & Error bar \\ 
        \hline
        0.001 & 0.063 & 0.044 & 0.084 & 0.030 \\ 
        \hline
        0.037& 0.090  & 0.037 & 0.101 & 0.029 \\ 
        \hline
        0.072 & 0.173  & 0.026 & 0.102 & 0.028 \\ 
        \hline
        0.108 & 0.139  & 0.019 & 0.085 & 0.024 \\ 
        \hline
        0.143 & 0.046  & 0.028 & 0.092 & 0.023 \\ 
        \hline
        0.179 & 0.064  & 0.040 & 0.080 & 0.020 \\ 
        \hline
        0.214 & 0.121  & 0.049 & 0.061 & 0.016 \\ 
        \hline
        0.226 & 0.054  & 0.059 & 0.054 & 0.016 \\ 
        \hline
        0.238 &0.108  & 0.049 & 0.004 & 0.013 \\ 
        \hline
        0.244& 0.088  & 0.024 & 0.037 & 0.010 \\ 
        \hline
       0.249 & 0.066  & -0.030 & 0.008 & 0.007 \\ 
        \hline
        \hline
    \end{tabular}
    \qquad\qquad\qquad
    \begin{tabular}{|c|c|c|c|c|}
       \hline
       \hline
       $ \tilde{\lambda} $ & $ \mathds{D}_{\rm tot} $ & $ \mathds{D}_{\rm pps} $ & $ \mathds{D}_{\rm deco} $ & Error bar \\ 
       \hline
       0.2 & -0.0334  & -0.006 & -0.048 & 0.006 \\ 
       \hline
       0.25&-0.0378 & -0.002 & -0.020 & 0.002 \\ 
       \hline
       0.3&-0.0085 & -0.004 & -0.018 & 0.001 \\ 
       \hline
       0.35&-0.0066   & -0.002 & -0.020 & 0.001 \\ 
       \hline
       0.4&0.0004 & -0.008 & -0.016 & 0.001 \\ 
       \hline
       0.45&-0.0018  & -0.003 & -0.008 & 0.001 \\ 
       \hline
       0.5&0.0097    & 0.000 & -0.004 & 0.001 \\ 
       \hline
       0.63&0.0083  & 0.000 & 0.001 & 0.001 \\ 
       \hline
       0.8&0.0130   & -0.001 & 0.002 & 0.001 \\ 
       \hline
       1&0.0097 & 0.001 & 0.003 & 0.001 \\ 
       \hline
       1.2&0.0089   & 0.001 & 0.004 & 0.001 \\ 
       \hline
       \hline
   \end{tabular}
    \label{errortable}
    \caption{The left table is the errors of ZPF (FIG.\,2 of the main text). The right table is the errors of the order parameters (FIG.\,3 of the main text), where $\Omega/\omega =50$. The $\mathds{D}_{\rm tot}$ column is the total error of ZPF (or the order parameter), which is defined as the deviation of experimental results from theory results. The 3rd and 4th columns corresponding to the errors influenced by the imperfect experimental PPS and decoherence, respectively. The last column represents the standard deviations of ZPF (or the order parameter) coming from the fitting errors. }\label{table}
\end{table}

\section{Experimental errors and antisqueezing enhanced SNR}
\begin{figure}[H]
    \centering
    \includegraphics{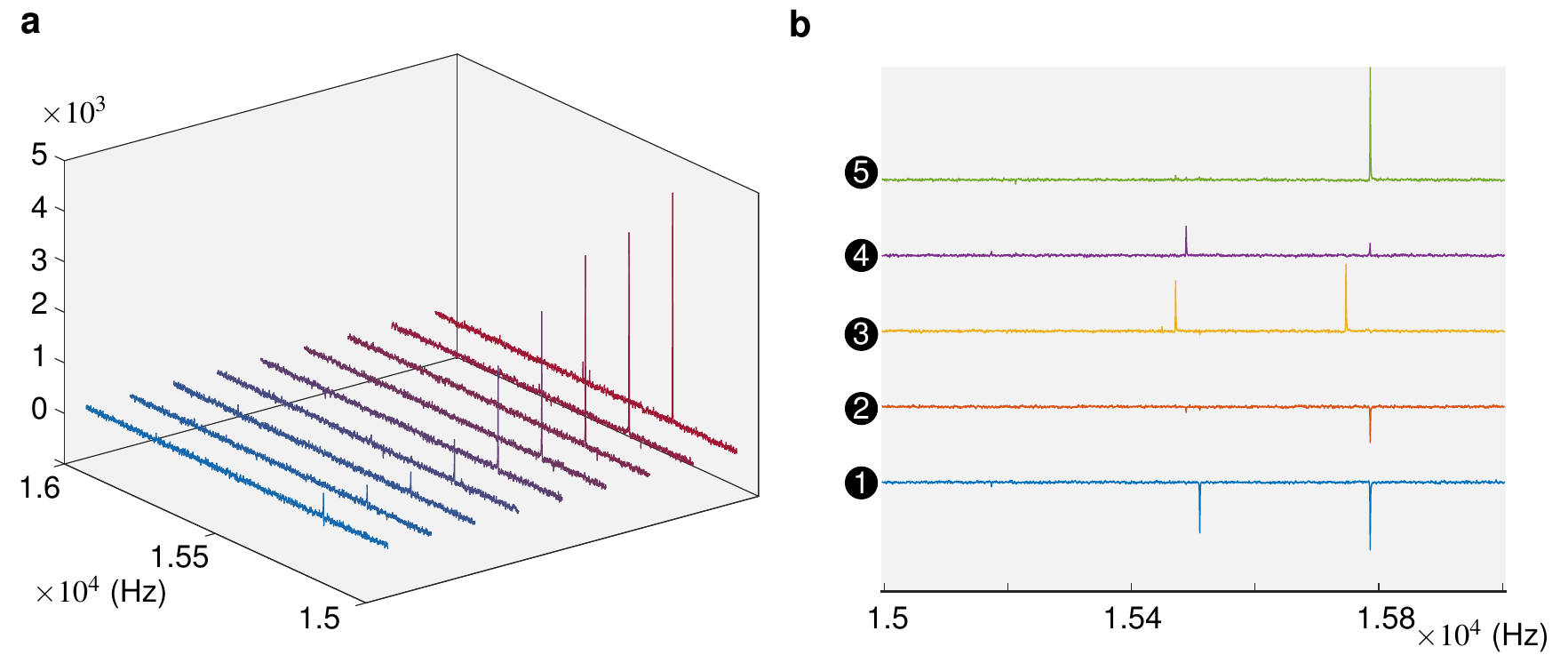}
    \caption{{\bf The NMR spectra.} {\bf a,} The readout spectra for obtaining the ZPF of system. As the antisqueezing effect increasing, the signal-to-noise rate increases. {\bf b,} Five NMR readout spectra for measuring order parameters of SPT. Spectra (1-3) are obtained by applying $\pi/2$ pulse along $y$ axis to three \textsuperscript{19}F spins. Spectrum 4 corresponds to read out the fourth qubit after applying operation $\hat{U}_1$ to the system. Reading out directly the fourth spin gives spectrum 5.}    
    \label{Supp-fig5}
\end{figure}

The errors in our experiment include the deviation of experimental datas from theoretical results and the variances of experimental data (i.e., errors bar). Different experimental errors on measuring the ZPF and order parameter are shown in the left and right parts of TABLE\,\ref{table}, respectively. Without loss the generality, we choose the case of measuring ZPF as example (i.e., the errors in FIG.\,2 of the main text) to discuss different experimental errors.  On the one hand, the deviation mainly comes from the decoherence and the initial state imperfection. Firstly, we numerically apply ideal squeezing operator to the experimentally reconstructed initial state $\hat{\rho}_{\rm pps}$, followed by an ideal readout. Then we get the ZPF value which is mainly influenced by the initial state imperfection. The deviations between the influenced ZPF and theoretical ZPF are shown in the third column of TABLE\,\ref{table}. Secondly, the effects of decoherence in NMR system can be described by the generalized amplitude damping channel $\epsilon_{GAD}$ and the phase damping channel $\epsilon_{PD}$~\cite{vandersypen2001experimental}. For small duration $\Delta t$, the errors of the phase damping channel is involved by $\hat{\rho} \rightarrow \epsilon_{PD}^{(4)} \circ  \epsilon_{PD}^{(3)} \circ\epsilon_{PD}^{(2)} \circ \epsilon_{PD}^{(1)} (\hat{\rho}) $, where $\epsilon_{PD}^{(i)} (\hat{\rho}) = (1-p_i) \hat{\rho} + p_i \hat{\sigma}_{z}^{(i)} \hat{\rho}\hat{\sigma}_{z}^{(i)}$, and $p_i = \frac{1}{2}\left[1- \exp(-{\Delta t}/{T^{(i)}_2})\right]~(i=1,2,3,4)$. Here $T^{(i)}_2$ is the decoherence parameter which can be found in FIG.\,1 of the main text. Similarly, the generalized amplitude damping error is characterized as $\hat{\rho} \rightarrow \epsilon_{GAD}^{(4)} \circ  \epsilon_{GAD}^{(3)} \circ\epsilon_{GAD}^{(2)} \circ \epsilon_{GAD}^{(1)} (\hat{\rho}) $ and it is calculated by $\epsilon_{GAD}^{(i)} (\hat{\rho}) = \sum _ {s} E^{(i)} _s {\hat{\rho} E^{(i)\dagger}_s}$, where
\begin{equation}
    \begin{aligned}
        E_{1}^{(i)}=\sqrt{p}
        \begin{pmatrix}
        1 & 0 \\
        0 & \sqrt{1-\eta^{(i)}}
        \end{pmatrix}
        , E_{2}^{(i)}=\sqrt{1-p}
        \begin{pmatrix}
            0 & 0 \\
        \sqrt{\eta^{(i)}} & 0
        \end{pmatrix},
         \\
        E_{3}^{(i)}=\sqrt{1-p}\left(\begin{array}{cc}
        \sqrt{1-\eta^{(i)}} & 0 \\
        0 & 1
        \end{array}\right), E_{4}^{(i)}=\sqrt{p}\left(\begin{array}{cc}
        0 & \sqrt{\eta^{(i)}} \\
        0 & 0
        \end{array}\right),
    \end{aligned}
\end{equation}
with $\eta^{(i)}=1-\exp(-\Delta t/T^{(i)}_1)$, $p\approx 1/2$, and $T^{(i)}_1$ is the decoherence parameter shown in FIG.\,1 of the main text. With ideal initial state and measurement, the numerical simulation shows that the decoherence has much influence on the final ZPF, as shown in the fourth column of TABLE\,\ref{table}. On the other hand, the readout errors shown in the last column of TABLE\,\ref{table} mainly come from the statistical fluctuation of the NMR spectra. In our experiment, the peak intensities are obtained by fitting the NMR spectra to a sum of Lorentz functions. The results of fitting indicate that these intensities have about 0.002 standard error, which will contribute to a standard error of the ZPF through error propagation formula. The variance errors of ZPF is
\begin{equation}
     \Delta_{\rm ZPF}^2= \frac{1}{4} \frac{1}{\rm ZPF^2} \Delta^2 (\tr(\hat{\rho} \hat{x}^2))    +   \frac{1}{\rm ZPF^2} \tr(\hat{\rho} \hat{x}) \Delta^2(\tr(\hat{\rho} \hat{x})),
     \label{Eq-delta_ZPF}
\end{equation}
where $\Delta^2 [\tr(\hat{\rho} \hat{x}^2)]$ and $\Delta^2[\tr(\hat{\rho} \hat{x})]$ are the variance errors of $\tr(\hat{\rho} \hat{x}^2)$ and $ \tr(\hat{\rho} \hat{x})$, respectively, which are obtained from the errors of peak intensities through linear error propagation. Interesting conclusion is obtained from Eq.\,(\ref{Eq-delta_ZPF}) and the last column of TABLE\,\ref{table}, i.e., the variance (or errors bar) of ZPF decrease as increasing the antisqueezing effect. This originally comes antisuqeezing enhanced signal-to-noise ratio (SNR) of the NMR spectra. As shown in FIG.\,\ref{Supp-fig5}(a), a typical NMR readout spectra shows that the signal peak increase as increasing the antisqueezing effect, whereas the noises almost keep a constant. This conclusion is consistent with FIG.\,2 of the main text. 
